\documentclass[twocolumn,english,aps,prl]{revtex4}
\pdfoutput=1
\usepackage[T1]{fontenc}
\usepackage[latin9]{inputenc}
\usepackage{color}
\usepackage{amsmath}
\usepackage{graphicx}
\usepackage{amssymb}
\usepackage{esint}
\usepackage{wasysym}

\makeatletter
\@ifundefined{textcolor}{}
{%
 \definecolor{BLACK}{gray}{0}
 \definecolor{WHITE}{gray}{1}
 \definecolor{RED}{rgb}{1,0,0}
 \definecolor{GREEN}{rgb}{0,1,0}
 \definecolor{BLUE}{rgb}{0,0,1}
 \definecolor{CYAN}{cmyk}{1,0,0,0}
 \definecolor{MAGENTA}{cmyk}{0,1,0,0}
 \definecolor{YELLOW}{cmyk}{0,0,1,0}
 }

\@ifundefined{definecolor}
 {\usepackage{color}}{}
\@ifundefined{definecolor}{\@ifundefined{definecolor}
 {\usepackage{color}}{}
}{}
\usepackage{epsf}

\newcommand{\lscoo}{La$_{5/3}$Sr$_{1/3}$CoO$_{4}$}
\newcommand{\lscop}{La$_{1.7}$Sr$_{0.3}$CoO$_{4}$}
\newcommand{\lscox}{La$_{2-x}$Sr$_{x}$CoO$_{4}$}

\newcommand{\w}{\omega}

\newcommand{\wR}{\omega_{\rm S}}
\newcommand{\Qc}{\vec{Q}_{\rm C}}
\newcommand{\Qs}{\vec{Q}_{\rm S}}
\newcommand{\xic}{\xi_{\rm C}}
\newcommand{\xicpa}{\xi_{\rm C}^\parallel}
\newcommand{\xicpe}{\xi_{\rm C}^\perp}
\newcommand{\xis}{\xi_{\rm S}}
\newcommand{\xispa}{\xi_{\rm S}^\parallel}
\newcommand{\xispe}{\xi_{\rm S}^\perp}
\newcommand{\Sc}{S_{\rm C}}
\newcommand{\Ss}{S_{\rm S}}

\newcommand{\TCO}{T_{\rm CO}}

\newcommand{\HSP}{\mathcal{H}_{\rm sp}}

\makeatother

\usepackage{babel}

\begin{document}

\title{
Disorder, cluster spin glass, and hourglass spectra in striped magnetic insulators
}

\author{Eric C. Andrade}
\author{Matthias Vojta}
\affiliation{Institut für Theoretische Physik, Technische Universit\"at Dresden,
01062 Dresden, Germany}


\begin{abstract}
Hourglass-shaped magnetic excitation spectra have been detected in a variety of doped
transition-metal oxides with stripe-like charge order. Compared to the predictions of
spin-wave theory for perfect stripes, these spectra display a different intensity
distribution and anomalous broadening. Here we show, based on a comprehensive modelling
for \lscoo, how quenched disorder in the charge sector causes frustration, and
consequently cluster-glass behavior at low temperatures, in the spin sector. This
spin-glass physics, which is insensitive to the detailed nature of the charge disorder,
but sensitive to the relative strength of the magnetic inter-stripe coupling, ultimately
determines the distribution of magnetic spectral weight: The excitation spectrum,
calculated using spin waves in finite disordered systems, is found to match in detail the
observed hourglass spectrum.
\end{abstract}

\date{\today}


\maketitle

Magnetic excitations, as measured by inelastic neutron scattering (INS), serve as a
fingerprint of the underlying magnetic state. In layered transition-metal oxides,
so-called {}``hourglass'' spectra are frequently observed
\cite{arai99,jt04,hayden04,buyers04,christensen04,hinkov07,reznik08c,xu09,booth11,braden11},
consisting of intensity at incommensurate (IC) wavevectors at low energy which, with
increasing energy, first disperses towards the antiferromagnetic (AF) wavevector
$(\pi,\pi)$, leading to a large-intensity {}``resonance'', and then disperses outwards
again. In the case of cuprate superconductors, explanations both in terms of
weak-coupling excitations of a Fermi liquid and of strong-coupling collective modes of an
underlying stripe state have been proposed \cite{eremin05b,vvk,hirschfeld10,kiv_rmp,ap}.

More recently, hourglass spectra have been detected in doped cobaltates and manganites
\cite{booth11,braden11}. Because these are insulators, the magnetic excitations can be
conclusively linked to collective local-moment physics. Moreover, charge order is known
to occur, i.e., magnetic and non-magnetic ions form a pattern of unidirectional
two-dimensional (2D) stripes \cite{braden12}. Indeed, assuming magnetic moments to be
arranged in stripes and coupled by superexchange interactions allows one to qualitatively
understand some gross features of the measured spectra.

However, the experimental data display pronounced deviations from
the spin-wave spectra of perfect stripes: (i) At low energies, outward
dispersing excitation branches are essentially completely missing.
(ii) The data typically show a large and anisotropic intrinsic broadening. (iii) For a
sizeable range of intermediate energies a ``vertical dispersion'' is observed near
$(\pi,\pi)$. (iv) Compared to perfect-stripe spin-wave results, spectral weight is
redistributed from higher energies into the broad ``resonance'' (or saddle-point) feature.
It has been suggested that quenched disorder plays a role in explaining these features,
but concrete calculations have been restricted to ad-hoc broadening of perfect-stripe results,
and deeper insights are lacking \cite{booth11,braden11,braden12}.

In this paper we analyse spin-wave spectra of stripes in 2D insulators, treating
disorder exactly in large finite-size systems. Importantly, stripe disorder generically
leads to magnetic frustration, because charge stripes act as antiphase domain walls for
the AF order. Our results identify the hourglass as the excitation spectrum of the
low-temperature cluster-spin-glass state which emerges due to the disorder-induced
frustration. We discuss the ingredients required for this spin-glass physics and the
implications for different material
families. 

\begin{figure}
\includegraphics[width=0.47\textwidth]{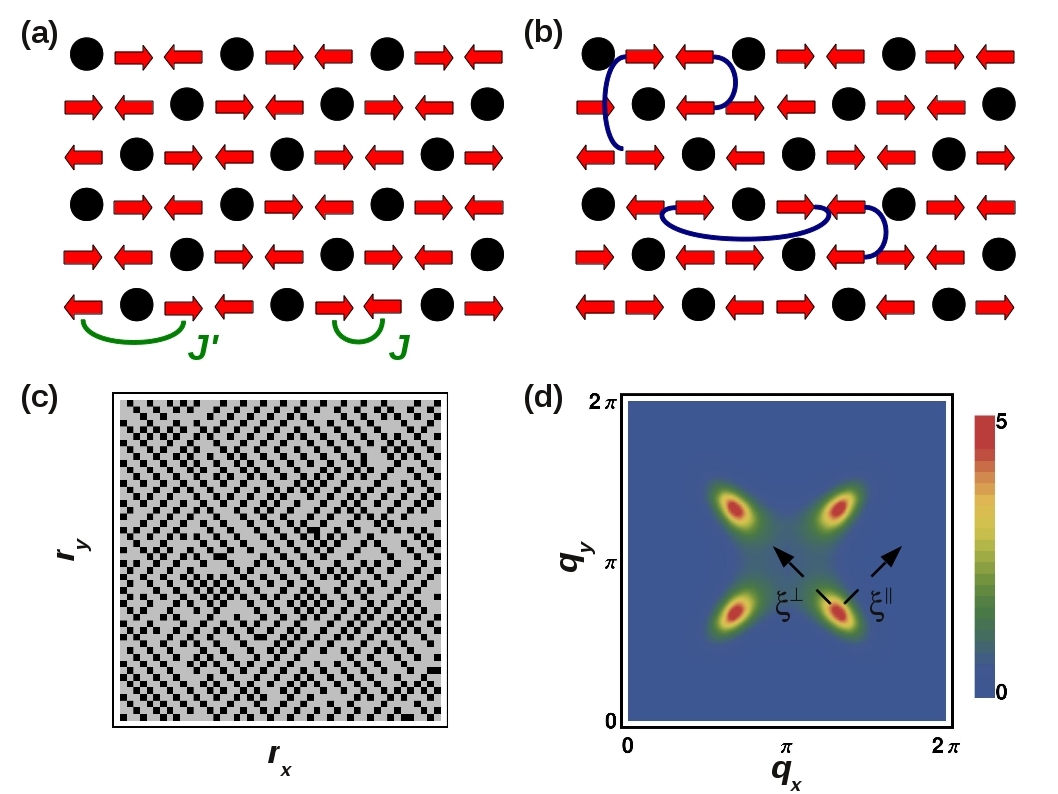}
\caption{
\label{fig:charge_dis}
(a) Perfect stripe order in the $\mbox{Co}\mbox{O}_{2}$
planes of \lscoo. Circles (arrows) represent non-magnetic Co$^{3+}$
(magnetic Co$^{2+}$) ions. Both magnetic couplings $J$ and $J'$
are AF. (b) Frustrated bonds caused by disorder in the charge sector.
(c) Disordered charge configuration with $\xicpa\approx5$ and $\xicpe\approx2$,
obtained from the Ising model \cite{suppl} for $L=48$. Black (gray)
squares correspond to Co$^{3+}$ (Co$^{2+}$) ions. (d) Charge structure
factor $\Sc(\vec{q})$ for the same parameters as in (c) averaged
over $10^{3}$ charge configurations. 
}

\end{figure}

\textit{\lscoo.}
For definiteness, we choose to model the magnetism
in \lscoo, for which comprehensive INS data is available \cite{booth11}.
\lscox\ shows signatures of stripe order for $0.3<x<0.6$ \cite{cwik09,braden12}.
For $x=1/3$ the Co$^{2+}$ and Co$^{3+}$ ions form a robust diagonal
period-3 stripe pattern, with charge order setting in at $\TCO$ far
above room temperature \cite{cwik09,CO_foot}. Co$^{2+}$ is in a $S=3/2$
high-spin state, while Co$^{3+}$ is in a $S=0$ low-spin state \cite{booth11,hollm08},
and from the properties of the $x\!=\!0$ \cite{babke10} and $x\!=\!0.5$
\cite{helme09} compounds it is known that the magnetic couplings
-- $J$ between neighboring Co$^{2+}$ and $J'$ across a non-magnetic
Co$^{3+}$ -- are both AF. For perfect charge order, this results
in a stripe pattern as depicted in Fig.~\ref{fig:charge_dis}(a),
exhibiting IC magnetism with magnetic Bragg peaks at $\Qs=(\pi\pm\pi/3,\pi\mp\pi/3)$.
{[}Stripes running along the opposite diagonal have $\Qs=(\pi\pm\pi/3,\pi\pm\pi/3)$.{]}


\textit{Disordered charge stripes.}
The experiments in Ref.~\onlinecite{booth11}
have proven substantial disorder effects to be present in \lscoo.
It is plausible to assume that those originate from static imperfections
in the charge order. In contrast, dynamic fluctuations of the charge
order are likely negligible for $T\ll\TCO$ because of the robust
insulating nature of the material which is re-enforced by spin-blockade
effects \cite{cwik09,braden12,maignan04,chang09}.

We therefore start by modelling stripes of Co$^{2+}$ and Co$^{3+}$ ions. We construct an
Ising model for variables $n_{i}$ on the sites $i$ of a square lattice, where $n_{i}=0$
refers to a Co$^{2+}$ and $n_{i}=1$ to a Co$^{3+}$ ion. The model is chosen such that it
has perfect stripe configurations with ordering wavevector
$\Qc=(\pi\pm\pi/3,\pi\mp\pi/3)$ {[}as in Fig. \ref{fig:charge_dis}(a){]} and
$\Qc=(\pi\pm\pi/3,\pi\pm\pi/3)$ as ground states at fixed filling $\langle n\rangle=1/3$,
for details see Ref.~\onlinecite{suppl}.

Monte-Carlo (MC) simulations of this model on $L\times L$ lattices ($L\leq 48$)
are used to generate charge configurations away from this ordered
state, i.e., with well-defined short-range order \cite{suppl}.
Representative results are in Fig. \ref{fig:charge_dis}(c) and Ref.~\onlinecite{suppl},
showing multiple stripe domains and various types of defects. 
We characterize these configurations by their correlation lengths
$\xicpa$ ($\xicpe$) parallel (perpendicular) to the stripe direction,
we obtain the $\xic$ from the charge structure factor $\Sc(\vec{q})$,
Fig. \ref{fig:charge_dis}(d), by a fit to a Lorentzian lineshape
near $\Qc$. 
We note that the procedure of generating disordered stripes is not
unique, as -- for fixed $\xic$ -- the types and distributions of
defects will depend on details of both the Ising model and the MC
protocol \cite{suppl}. However, such details have remarkably
little influence on the behavior of the spin sector.


%

\textit{Spin model.}
To model the magnetism of \lscoo, we place localized $S=1/2$ spins on the Co$^{2+}$
sites of a 2D disordered stripe configuration as in Fig.~\ref{fig:charge_dis}(b,c). The
spins are assumed to interact via Heisenberg exchanges as in
Fig.~\ref{fig:charge_dis}(a):
\begin{equation}
\HSP=\sum_{i,j}\sum_{\alpha}J_{ij}^{\alpha}S_{i}^{\alpha}S_{j}^{\alpha}.
\label{eq:h_spin}
\end{equation}
The first sum runs over the lattice sites with $n_{i}=0$ according
to a given charge configuration $\{n_{i}\}$, and $\alpha=x,y,z$.
Guided by Refs.~\onlinecite{booth11,helme09,babke10} we assume that the
nearest-neighbor coupling $J$ (which sets our energy scale) has the same form as in the
undoped parent compound: $J^{x}=J\left(1+\epsilon\right)$, $J^{y}=J$, and
$J^{z}=J\left(1-\delta\right)$. The parameters $\delta$ and $\epsilon$ control the spin
anisotropy with $\delta=0.28$ and $\epsilon=0.013$ \cite{helme09,babke10,suppl}. Because the
coupling across a Co$^{3+}$ ion, $J'$, is considerably smaller than $J$, we assume it to be
isotropic with typically $J'=0.05J$. Interlayer couplings are neglected.


%
\begin{figure*}
\includegraphics[width=1.02\textwidth]{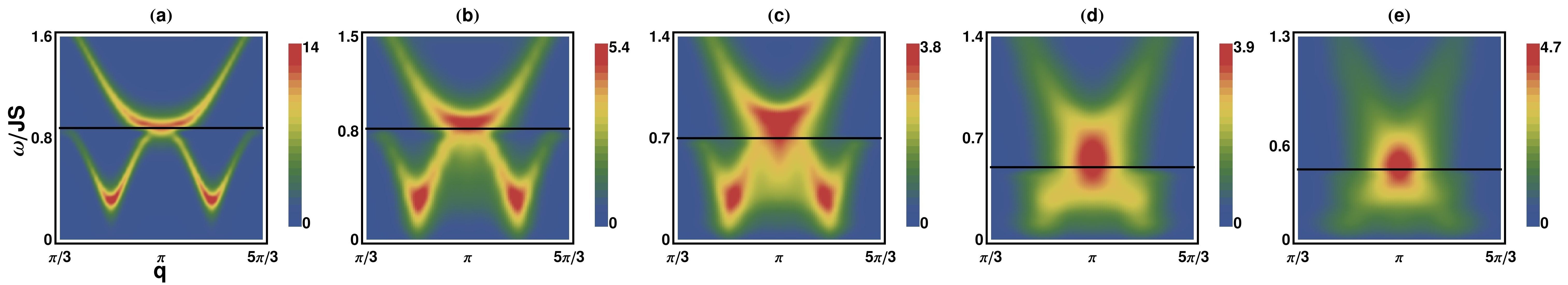}
\caption{\label{fig:hour1}
Magnetic excitations, $\chi''(\vec{q},\w)$, of
disordered stripes. From (a) to (e) the disorder in the charge sector
increases, as characterized by the correlation lengths $\xicpa$,
$\xicpe$. Each panel is divided at $\w=\wR$, with the lower {[}upper{]}
portion showing data along $\vec{q}=(q,q)$ {[}$(q,\pi)${]} in order
to represent the high-intensity features. Parameters are $J'=0.05J$,
$L=36$, and (a) Perfect stripes, $\wR=0.88JS$; (b) $\xicpa\simeq19$,
$\xicpe\simeq8$, $\wR=0.82JS$; (c) $\xicpa\simeq10$, $\xicpe\simeq4$,
$\wR=0.70JS$; (d) $\xicpa\simeq5$, $\xicpe\simeq2$, $\wR=0.50JS$;
(e) $\xicpa\simeq3$, $\xicpe\lesssim1$, $\wR=0.45JS$ \cite{suppl}.
The $\delta$ peaks in Eq.~\eqref{eq:chi_sw} were replaced by Lorentzians
with width $\Gamma=0.1JS$.}

\end{figure*}

\textit{Spin-glass ground state and excitations.}
For imperfect charge
order, $\HSP$ describes frustrated magnetism: As illustrated in Fig.~\ref{fig:charge_dis}(b),
stripe defects generically lead to frustrated couplings between neighboring
AF domains. This combination of disorder and frustration is expected
to lead to spin-glass behavior, more precisely {\em cluster spin-glass}
behavior, because the entities subject to frustrated couplings are
not single spins, but domains of varying size. This spin glass displays
short-range magnetic order (inherited from the perfect stripe state),
with magnetic correlation lengths $\xispa$, $\xispe$ which we determine
from the spin structure factor $\Ss(\vec{q})$. 
In general, we find that $\xis$ follows $\xic$ \cite{suppl}, but
for a strongly disordered charge sector with $\xicpa\lesssim4$, $\Ss(\vec{q})$
is no longer peaked at the IC $\Qs$, but instead displays a broad
maximum at $(\pi,\pi)$. This implies that the real-space period-3
structure is lost, and a spin glass with very short-ranged AF correlations
emerges.

To access the spin dynamics, we first determine locally stable classical
states of $\HSP$ for a given charge configuration
$\{n_{i}\}$: For classical spins $\vec{S}_{i}$, we perform MC simulations
of $\HSP$ utilizing single-site updates, combining the heat-bath
and microcanonical (or over-relaxation) methods. To efficiently sample
all spin configurations, 
we also employ the parallel-tempering algorithm \cite{mc_ea}. 
This procedure is followed by a quench down to $T=0$ using a greedy
conjugate-gradient algorithm. Our simulations confirm spin-glass behavior,
as detailed in Ref.~\onlinecite{suppl}.

We then calculate the excitation spectrum of $\HSP$ using linear spin-wave theory on
finite lattices: Deviations from a classical state are represented by non-interacting
bosons \cite{suppl} modelling Gaussian magnetic fluctuations.
The dynamic susceptibility is given by:
\begin{equation}
\chi''(\vec{q},\w)=\big[\sum_{\alpha}\sum_{\nu}\left|\left\langle 0\left|S^{\alpha}\left(\vec{q}\right)\right|\nu\right\rangle \right|^{2}\delta\left(\w-\w_{\nu}\right)\big]_{av},\label{eq:chi_sw}
\end{equation}
where $S^{\alpha}(\vec{q})$ is the Fourier-transformed spin operator,
$\left|0\right\rangle $ the magnon vacuum, and $\left|\nu\right\rangle $
a single-magnon state with energy $\w_{\nu}$. We typically evaluate
the average $\left[\cdots\right]_{av}$ over $80$ spin states obtained
from $40$ charge configurations. Formally, such a calculation yields
the zero-temperature excitation spectrum of a system with spatially
disordered static stripes, but the results continue to apply at finite
temperatures $T$ and even for slowly fluctuating stripes, provided
that both temperature and fluctuation frequency are small compared
to the energies $\w$ of interest in $\chi''(\vec{q},\w)$.



\textit{Charge disorder and hourglass spectrum.}
Fig.~\ref{fig:hour1} shows our central result, namely the evolution of
$\chi''(\vec{q},\w)$ with varying degree of stripe disorder for weak inter-stripe
coupling (i.e. large $J/J'$). The perfect-stripe case Fig.~\ref{fig:hour1}(a) has sharp
magnon peaks; at low energies gapped magnon branches (recall the magnetic anisotropy)
disperse both inward and outward from $\Qs$, and a sharply peaked large intensity maximum
occurs at $(\pi,\pi)$ corresponding to a saddle point of the spin-wave dispersion.
In contrast, as disorder increases, Figs.~\ref{fig:hour1}(b,c), the excitations are
broadened both in momentum and energy. The anisotropy gap is smeared, and the intensity at
low energies is enhanced as compared to the clean case: Disorder causes a strong redistribution
of spectral weight from high to low energies \cite{suppl}.
Even more strikingly, for the strongly disordered case in Figs.~\ref{fig:hour1}(d,e), the
outward dispersing branches at low energy have disappeared completely. Instead, the
weight is concentrated near $(\pi,\pi)$ over a broad energy range, leading to an apparent
{}``vertical'' dispersion. Finally, in Fig.~\ref{fig:hour1}(e) the charge disorder is so
strong that the IC signal at small $\w$ is hardly visible; here,
$\Ss\left(\vec{q}\right)$ shows a broad maximum at $(\pi,\pi)$ \cite{suppl}.

A remarkable trend is that disorder causes -- apart from broadening -- a strong downward
renormalization of the saddle-point energy: For otherwise identical parameters, $\wR$ in
Fig.~\ref{fig:hour1}(e) is half of that in the perfect-stripe case.


\textit{Influence of model parameters.}
In the spin sector, the only parameters are the exchange constants $J$ and $J'$ and their
anisotropies, whose values we fixed based on experimental data on \lscoo\ \cite{booth11}
and its parent compound \cite{babke10}. Some uncertainty concerns the strength of the
inter-stripe coupling, i.e., the ratio $J/J'$. In fact, this ratio varies
considerably between different families of striped oxides, with values of 2 reported for
nickelates (which do not show an hourglass dispersion) \cite{booth03,woo05} and 7-20 for
cobaltates \cite{booth11}. Spin-wave calculations in the disorder-free case
\cite{booth03,yaoprb06,booth11} show that large $J/J'$ is crucial to the observation of
an hourglass-like spectrum, whereas for small $J/J'$ the saddle point moves to the top of
the spin-wave dispersion.
We have performed calculations for disordered stripes with varying
$J/J'$ which confirmed this analysis, for details see Ref.~\onlinecite{suppl}.

\begin{figure}[b]
\includegraphics[width=0.45\textwidth]{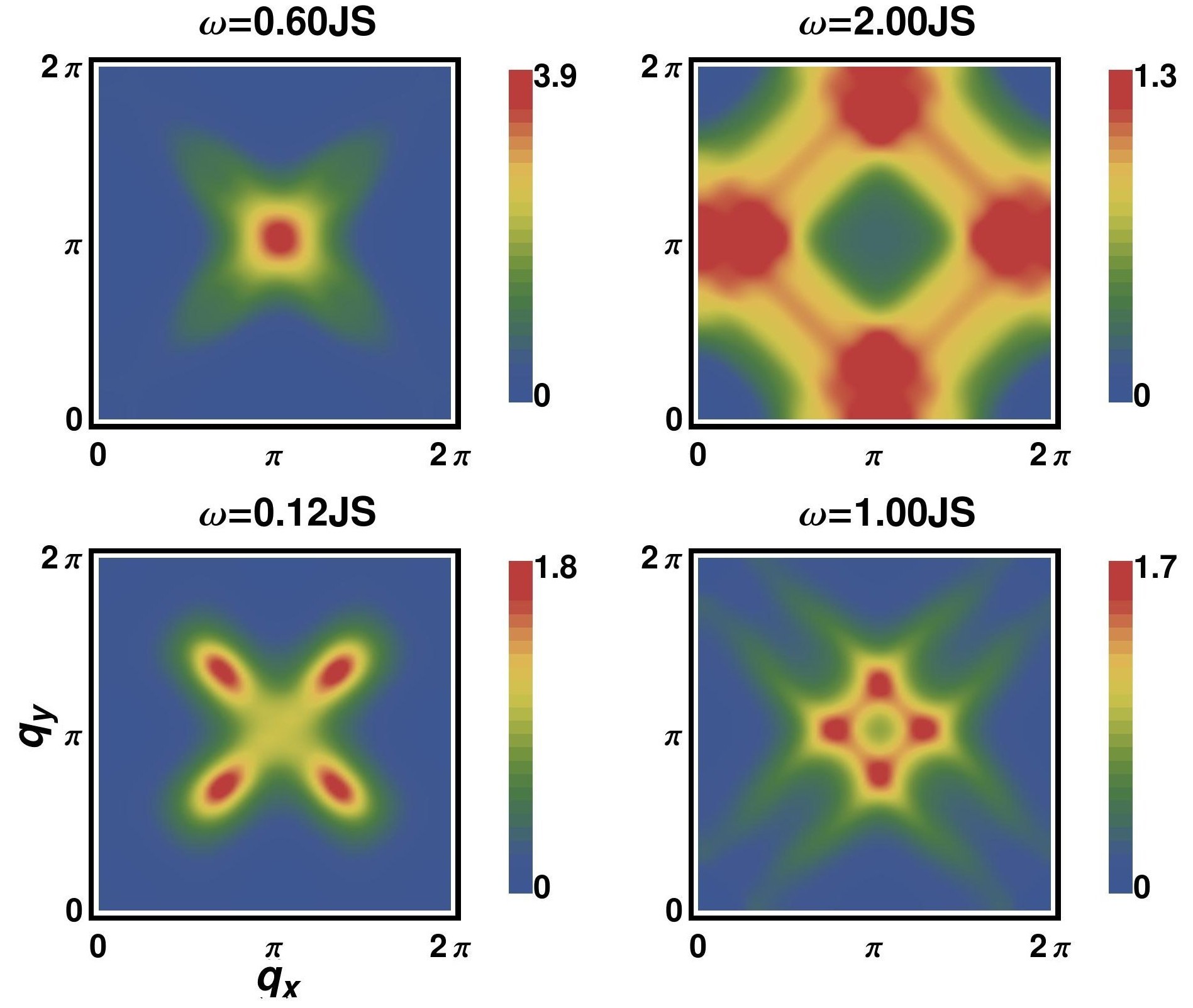}
\caption{\label{fig:ecut1}
Constant-energy cuts of $\chi''(\vec{q},\w)$ with model parameters as in
Fig.~\ref{fig:hour1}(d): $J'=0.05J$, $\xicpa\simeq5$, $\xicpe\simeq2$, $L=36$.
}
\end{figure}

Turning to the charge sector, our modelling here relies on a set of parameters and
assumptions \cite{suppl}, with rather little experimental guidance available
\cite{CO_foot}. We have tested different schemes to introduce stripe disorder, in
particular resulting in a varying degree of anisotropy $\xicpa/\xicpe$. We have found that
this has rather little influence on the magnetic behavior, such that our conclusions are
independent of details of the stripe disorder \cite{suppl}, and the only important
parameter is $\xic$.



\textit{Application to \lscox.}
A comparison of our results with the INS data of Ref.~\onlinecite{booth11} indicates that
charge disorder as in Fig.~\ref{fig:hour1}(d) is realized in \lscoo. For these
parameters, we show constant-energy cuts of $\chi''(\vec{q},\w)$ in Fig.~\ref{fig:ecut1}.
We notice that the low-energy inelastic response ($\w=0.12JS$) displays an
anisotropic broadening -- similar to the broadening of the Bragg signal,
Fig.~S5(a) \cite{suppl}, where $\xispa/\xispe\approx2.3$ -- a behavior which is generic
for our model. Significant low-energy intensity occurs also at
$(\pi,\pi)$, a feature that, in our calculations, is generically linked to an hourglass
spectrum.

\begin{figure}[t]
\includegraphics[width=0.49\textwidth]{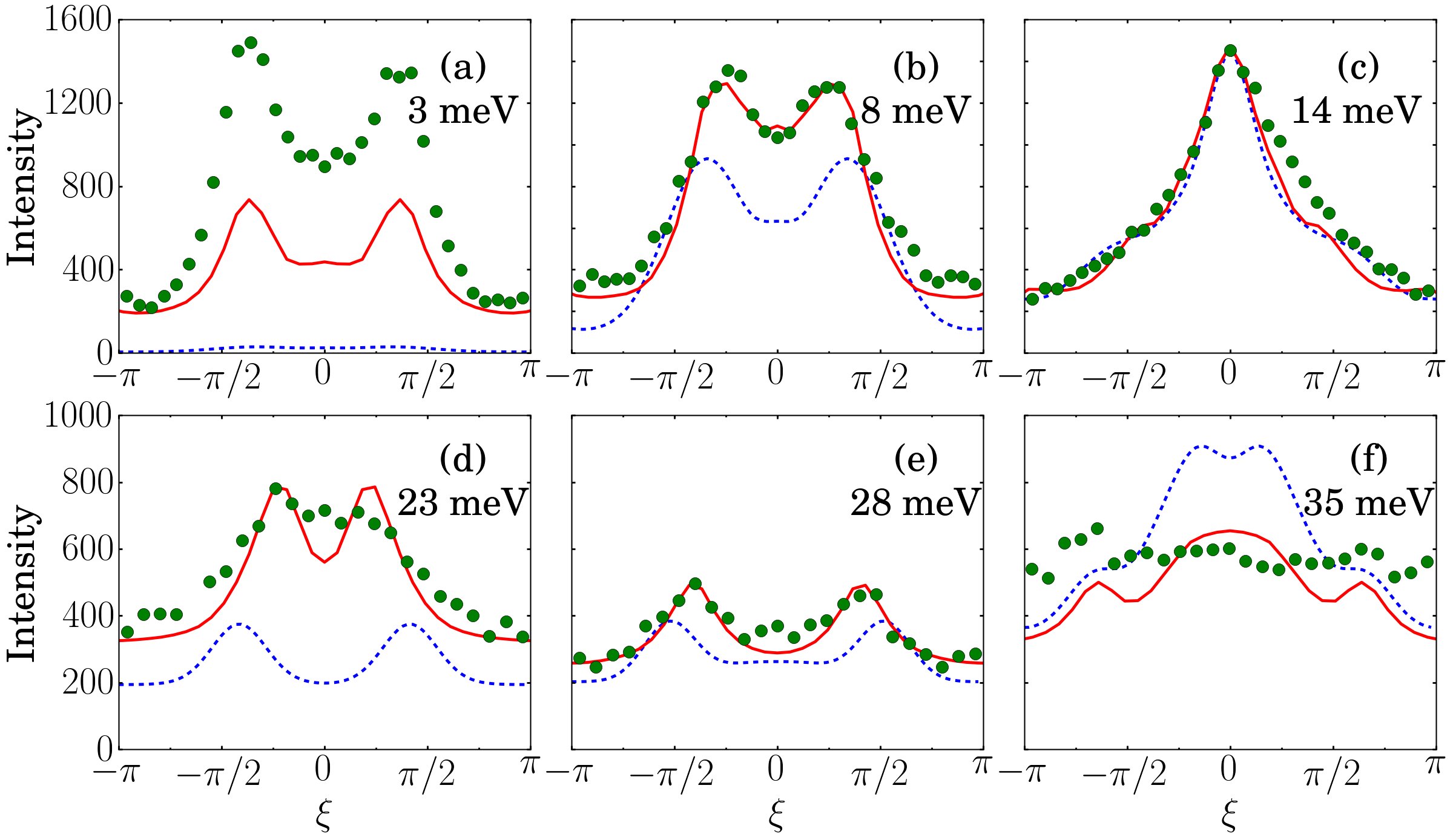}
\caption{\label{fig:comp}
Comparison of $\chi''(\vec{q},\w)$ for parameters as in Fig.~\ref{fig:hour1}(d) and
$JS=25$\,meV (solid) to the \lscoo\ data (dots) taken from Fig.~3 of
Ref.~\onlinecite{booth11}.
Panels (a-c): $\vec{q}=(\pi+\xi,\pi-\xi)$; (d-f): $\vec{q}=(\pi,\pi+\xi)$.
Also shown are ad-hoc broadened perfect-stripe spectra (dashed) following
Refs.~\onlinecite{booth11,suppl}, with $JS=18$\,meV adjusted to match the energy $\wR$ of the
saddle point.
For both theories, the energy broadening was $\Gamma=1$\,meV, and an overall intensity
factor and a constant background were adjusted by fitting the 14\,meV data and then were
used for all $\w$ \cite{int_note}. }
\end{figure}

For \lscoo, we deduce $JS\approx25$\,meV, for which we show an explicit comparison with
the experimental data in Fig.~\ref{fig:comp}. We find excellent agreement, apart from an
intensity mismatch at 3\,meV which is likely due to the simplified treatment of spin
anisotropies in our spin-1/2 model \cite{suppl}. Fig.~\ref{fig:comp} also shows the
result of ad-hoc momentum-space broadening of perfect-stripe spectra \cite{suppl}, as used in previous
modelling of the data \cite{booth11}. The most obvious failure of this approach concerns
the missing intensity at energies below and slightly above the saddle point; it also
overestimates the intensity of the upper spin-wave band which becomes visible near
$(\pi,\pi)$ at 35\,meV.
In addition, the mismatch of peak positions at elevated energies,
Fig.~\ref{fig:comp}(d,e),
illustrates that the real hourglass has approximately linearly dispersing branches above
the saddle point, while the perfect-stripe case yields a parabolic dispersion -- a
difference reproduced by our theory which becomes clear upon comparing panels (a) and (d)
of Fig.~\ref{fig:hour1}.

We conclude that the charge stripes in \lscoo\ are rather disordered, with an estimated
correlation length of $\xicpa\approx5$; this explains why charge order has not been
observed in scattering experiments. Our analysis predicts the magnetism in \lscoo\ to be
that of a cluster spin glass. Indeed, initial signatures for glassy behavior have been
detected \cite{booth_priv}, and a more thorough characterization is called for.

Finally, we propose that disordered stripes are also present in \lscox\ at $x\leq0.3$:
Smaller $x$ implies larger stripe spacing, with increased sensitivity to quenched
disorder. Indeed, both a broad Bragg signal \cite{cwik09} and broad low-energy
excitations \cite{cwikdiss} have been detected in \lscop\ around $(\pi,\pi)$, consistent
with the magnetism of strongly disordered stripes.


\textit{Other transition-metal oxides.}
For overdoped single-layer manganites with ordering wavevector
$\Qs=\left(2\pi\pm\delta,\pi\pm\delta\right)$ with $\delta=0.166\pi$, INS measurements
have recently detected hourglass spectra, in particular at elevated temperatures (while
low-$T$ spectra displayed outward dispersing branches) \cite{braden11}. The hourglass has
been linked to a short magnetic correlation length in this regime. Our model calculations
qualitatively confirm this interpretation: For intermediate $J/J'$, the coupling between
AF domains is robust at low $T$, but weakened at elevated $T$, such that the advertised
hourglass excitation spectrum of weakly correlated AF clusters is observed.

Finally, it is interesting to link our study to cuprate superconductors.
In fact, the physics of disordered stripes, either slowly fluctuating
or disorder-pinned, has been invoked to explain a variety of experimental
observations including INS data \cite{kiv_rmp,ap}. The hourglass
spectra observed in cuprates show a clear absence of outward dispersing
low-energy branches, and moreover display a pronounced piece of vertical
dispersion in the so-called pseudogap regime \cite{ap,jt04,hayden04,hinkov07}.
Assuming that our modelling qualitatively applies to cuprates as well,
we can link these experimental results to strongly disordered stripes.
We also note that very underdoped (but still superconducting) cuprates
have been found to display a quasieleastic ``central mode'' at $(\pi,\pi)$
which has been interpreted in terms of glassy short-range order \cite{buyers06}.
Considering that decreasing the doping level in cuprates below 1/8
weakens stripe order (such that it is more susceptible to quenched
disorder), this nicely matches our results: For strong stripe disorder,
$\Ss(\vec{q})$ looses its IC structure and instead displays
a broad peak at $(\pi,\pi)$ \cite{suppl}.


\textit{Summary.}
Our detailed modelling of disordered stripes in \lscoo\ reveals a link between the
celebrated hourglass excitation spectrum and a cluster-spin-glass magnetic ground state,
caused by frustrated couplings between locally ordered AF domains.
These phenomena appear common to a broad class of doped oxides.


We thank A. Boothroyd, M. Braden, and L. H. Tjeng for illuminating discussions,
and A. Boothroyd for providing the data of Ref.~\onlinecite{booth11}.
This research was supported by the DFG through FOR 960 and GRK 1621.




\begin{thebibliography}{99}
\bibitem{int_note}
The experimental data for $\w>14$\,meV in Fig.~3 of Ref.~\onlinecite{booth11} were taken
under conditions different from that for the low-$\w$ data, such that the intensity
scales are not strictly comparable \cite{booth_priv}.
\end{thebibliography}
\end{document}


\title{
Supplementary information for:\\
``Disorder, cluster spin glass, and hourglass spectra in striped magnetic insulators''
}

\author{Eric C. Andrade}
\author{Matthias Vojta}
\affiliation{Institut f\"ur Theoretische Physik, Technische Universit\"at Dresden,
01062 Dresden, Germany}

\date{\today}
\maketitle


\section{Generating charge configurations}
\label{sec:charge}

The charge order of Co$^{2+}$ and Co$^{3+}$ ions governs the magnetism
of \lscox. As this charge order sets in above room temperature, we
assume the charge order to be static on the time scale of magnetic
fluctuations, such that the spin sector can be analyzed -- in the
spirit of a Born-Oppenheimer approximation -- for a set of frozen
charge configurations. Those configurations correspond to imperfect
stripes, with imperfections likely caused by crystalline defects.

\begin{figure}
\includegraphics[width=0.43\textwidth]{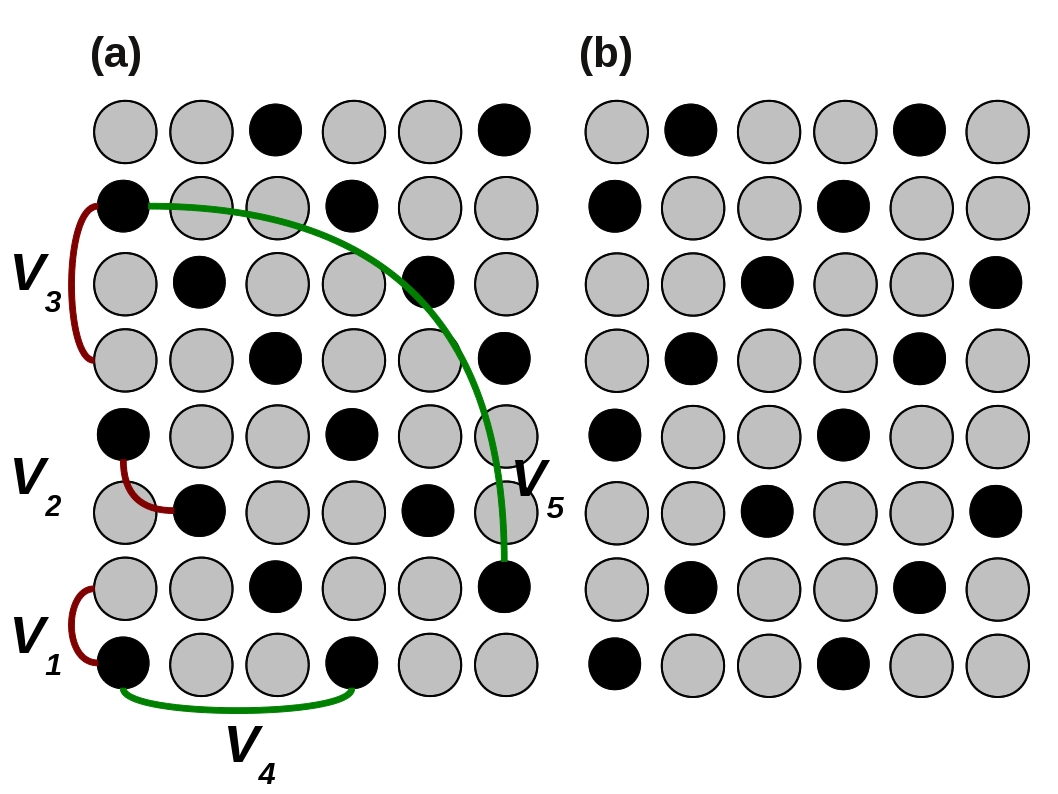}
\caption{\label{fig:charge_ideal}
Perfect diagonal period-3 stripe patterns
in the $\mbox{Co}\mbox{O}_{2}$ planes of \lscoo, where black and
gray circles correspond to $n_{i}=1$ (Co$^{3+}$) and $n_{i}=0$
(Co$^{2+}$), respectively. These configurations are ground states
of the Ising model in Eq.~\eqref{eq:h_charge}, with density-density
interactions $V_{1}>0$, $V_{2}<0$ and $V_{3}>0$ as shown in panel
a). The longer-range interactions $V_{4,5}$ will be utilized to modify
the structure of charge-order defects away from the ground state.
}
\end{figure}

We obtain such charge configurations from numerical simulations of
an appropriate Ising model. Consider variables $n_{i}$ on the sites
$i$ of a square lattice, where $n_{i}=0$ refers to a Co$^{2+}$ ion, while
$n_{i}=0$ refers to a Co$^{3+}$ ion. The $n_{i}$ are subject to a Hamiltonian of the form
\begin{eqnarray}
\mathcal{H}_{C} & = & \sum_{m=1}^{5}V_{m}\sum_{\left\langle i,j\right\rangle_{m}}n_{i}n_{j}
\label{eq:h_charge}
\end{eqnarray}
with a particle density of $\sum_{i}n_{i}=N/3$ adequate to describe
the compound \lscoo, where $N=L\times L$ is the number of sites, and periodic boundary
conditions are applied.
The $V_{m}$ represent density--density interactions as depicted in
Fig.~\ref{fig:charge_ideal}, which are chosen such that the ground
state of $\mathcal{H}_{C}$ is given by the diagonal stripe state established
for \lscoo, Fig.~\ref{fig:charge_ideal}. Specifically, $V_{1}$
is a nearest-neighbor repulsion, $V_{2}<0$ a second-neighbor attraction,
and $V_{3}>0$ a third-neighbor repulsion, while $V_{4,5}$ are longer-range
interaction used to modify the defect structure (see below, for most
of the calculations $V_{4,5}=0$).

We generate short-range-ordered charge configurations using Monte-Carlo (MC) simulations
of the Ising model in Eq.~\eqref{eq:h_charge}. As noted above, the relevant charge
disorder is {\em not} generated by thermal fluctuations at temperatures near or above a
putative charge-ordering transition (because this temperature is far above room
temperature in \lscox), but instead arises from structural defects (e.g. the dopant
distribution). As details of this quenched disorder are not known, and in order to keep
the set of input parameters minimal, we choose to employ the model
Eq.~\eqref{eq:h_charge} without quenched disorder, but instead extract configurations
away from thermal equilibrium.

To this end, the variables $\{n_{i}\}$ are initialized by randomly
distributing $N/3$ charges on $N$ sites. One MC step consists of
randomly selecting two sites and exchanging their charges, accepting
or rejecting this move according to the rules of the standard Metropolis
algorithm at a finite effective temperature $T$. We work at a low
temperature corresponding to the charge-ordered phase and select configurations
$\{n_{i}\}$ after a fixed number $\NMC$ of MC steps. Thus, for large
$\NMC$ the system reaches a state with long-range order, while smaller
$\NMC$ result in short-range-ordered states. We assume that these
states sensibly represent the situation in \lsco, because working
at low $T$ ensures to have well-ordered regions separated by localized
defect structures, as can be expected for a material far below its
charge-ordering temperature. Observables are finally obtained by averaging
over a large number of charge configurations, all obtained from runs
with different initial conditions -- this mimics the average over
a large disordered sample.

\begin{figure}
\includegraphics[width=0.45\textwidth]{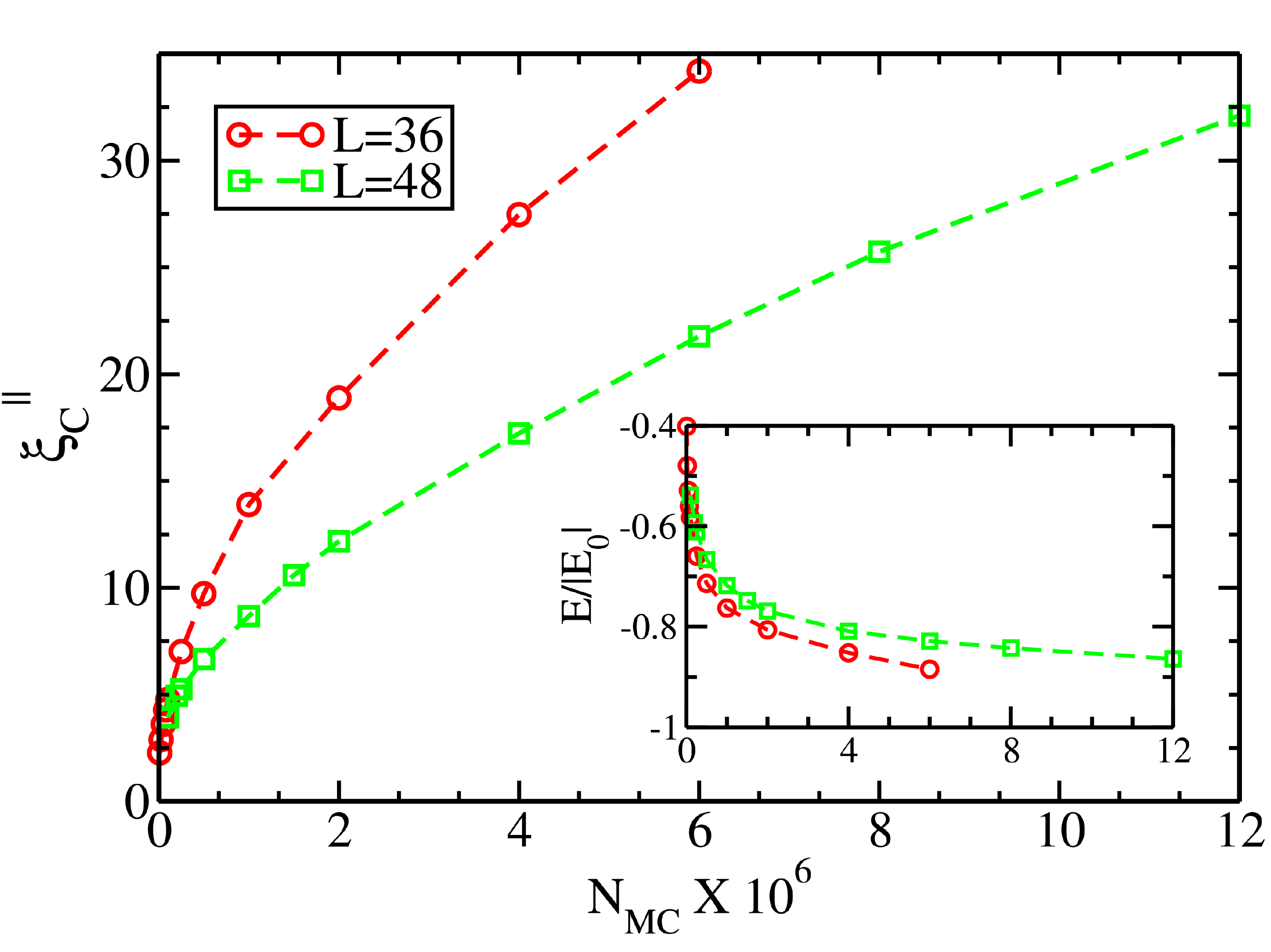}
\caption{\label{fig:xivsN}
Evolution of the charge correlation length $\xic$
with $\NMC$, i.e., along the MC simulation of the Ising model Eq.~\eqref{eq:h_charge}
for two different system sizes $L$. The parameters $V_{1}=4$, $V_{2}=-2$,
$V_{3}=4$, $V_{4,5}=0$ and $T=1$ correspond to the charge-ordered
phase, such that $\xic$ reaches the system size for large $\NMC$.
The inset shows the evolution of the energy $E=[\mathcal{H}_{C}]_{av}$
along the MC simulation, normalized to the energy of the perfectly
ordered state $E_{0}$.
}
\end{figure}

To quantify the charge disorder, we calculate the charge correlation
length $\xic$, which we obtain from the static charge structure factor
\begin{eqnarray}
\Sc\left(\vec{q}\right) & = & \frac{1}{\NS}\big[\sum_{i,j}\left(1-n_{i}\right)\left(1-n_{j}\right)e^{-i\vec{q}\cdot\vec{r}_{ij}}\big]_{av},\label{eq:sfac_charge}\end{eqnarray}
 where $\NS=2N/3$ is the number of Co$^{2+}$ ions (i.e. spins),
and $\left[\cdots\right]_{av}$ denotes the average over charge configurations.
For short-range order, $\Sc\left(\vec{q}\right)$ is typically peaked
at four wavevectors $\Qc$ (see Fig.~1(d) of the main paper and Figs.~\ref{fig:all1},
\ref{fig:all2}): the peaks at $\Qc=(2\pi/3,2\pi/3)$ and $\Qc=(4\pi/3,4\pi/3)$
correspond to stripe domains like those in Fig.~\ref{fig:charge_ideal}(a),
while peaks at $\Qc=(2\pi/3,4\pi/3)$ and $\Qc=(4\pi/3,2\pi/3)$ belong
to stripes running along the opposite diagonal, Fig. \ref{fig:charge_ideal}(b).
The width of the peaks is anisotropic, thus we distinguish correlation
lengths $\xicpa$ and $\xicpe$ parallel and perpendicular to the
stripe direction, respectively. The evolution of $\xic$ with the
number of steps, $\NMC$, within our non-equilibrium MC procedure
is shown in Fig.~\ref{fig:xivsN}. As anticipated, the ordered state
(with $\xic\gtrsim L$) is reached for large $\NMC$, while smaller
values, $\NMC=10^{4}$--$10^{6}$, are suitable for obtaining charge
configurations with $\xic\approx2$--$10$ lattice spacings.

Individual charge configuration from our MC simulations are shown
in the top rows of Figs.~\ref{fig:all1}(b)-(e) and \ref{fig:all2}. We
see that the protocol generates finite-size stripe domains of both
orientations, as intended. For a large range of parameters with $V_{4,5}=0$
as in Fig.~\ref{fig:all1} we generically find $\xicpa>\xicpe$,
i.e., the diagonal arrangement of neighboring charges tends to be
more stable than the spacing between the diagonal charge lines. $\Sc\left(\vec{q}\right)$
also reveals that for very large disorder, Fig. \ref{fig:all1}(e),
significant weight appears away from $\Qc$. This effect is even more
pronounced in the spin sector, as will be discussed below.

Charge configurations with $\xicpa<\xicpe$ can be generated on purpose,
e.g., by choosing the longer-range couplings $V_{4,5}\neq0$, which
stabilize the stripe spacing and cut charge lines, respectively. Results
are shown in Fig.~\ref{fig:all2}, which now show a visibly different
defect structure in the charge sector. Remarkably, the behavior on
the spin sector is rather insensitive to this change (see below).

\begin{figure}
\includegraphics[width=0.45\textwidth]{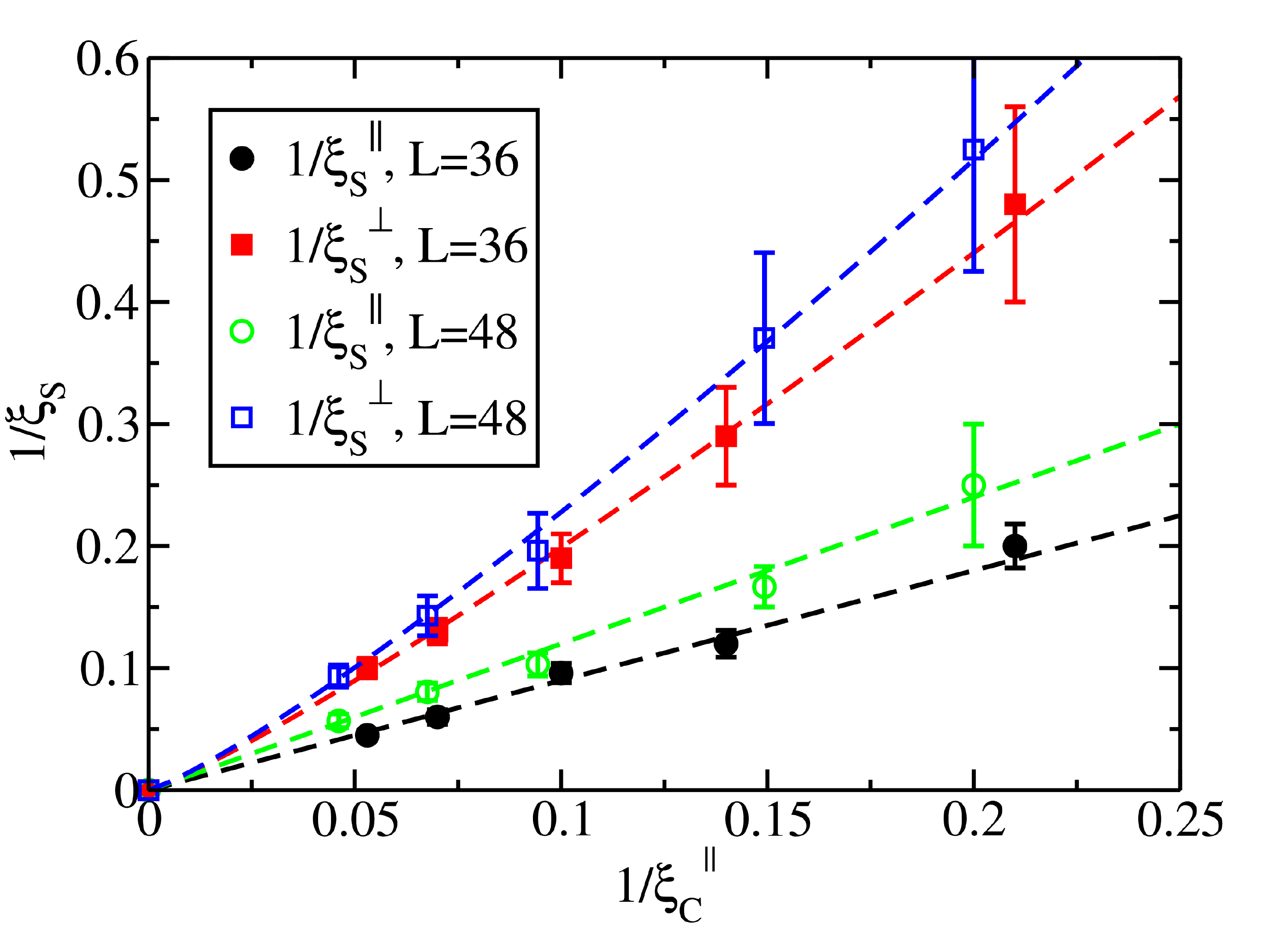}
\caption{
\label{fig:xis}Spin correlation lengths, $\xispe$ and $\xispa$,
of the classical ground state of $\HSP$, Eq.~\eqref{eq:h_spin},
as function of the charge correlation length $\xicpa$ which characterized
the stripe disorder. Clearly, $\xispa\approx\xicpa$ while $\xispe$
is smaller by a factor of $2$--$2.5$. }

\end{figure}


\section{Spin model}

As explained in the main text, the magnetism of \lscoo\ is assumed
to be well described by immobile magnetic moments that are located
on the Co$^{2+}$ sites (i.e. those with $n_{i}=0$) of a disordered
2D stripe configuration.

The Co$^{2+}$ ions are in a 3d$^{7}$ high-spin configuration with
$S=3/2$. A full modelling of these high-spin ions and their interactions
has been put forward in Refs.~\onlinecite{babke10,helme09} --
this involves spin and orbital angular momentum, crystal-field and
spin-orbit effects, as well as (nearly isotropic) exchange interactions.
As the local Hilbert space in this model contains 28 states, it is
more efficient to employ a reduced effective model instead: As shown
in Refs.~\onlinecite{babke10,helme09}, the low-energy part of
the excitation spectrum of \lsco\ both at $x=0$ and $x=0.5$ can
be well captured by a 2D spin-1/2 spin-only model.

Consequently, we adopt such a model for our purpose,
\begin{equation}
\HSP=\sum_{i,j}\sum_{\alpha}J_{ij}^{\alpha}S_{i}^{\alpha}S_{j}^{\alpha},
\label{eq:h_spin}
\end{equation}
and choose the nearest-neighbor $J$ as in Refs.~\onlinecite{babke10,helme09} to be
$J^{x}=J\left(1+\epsilon\right)$, $J^{y}=J$, and $J^{z}=J\left(1-\delta\right)$, with
$\delta=0.28$ and $\epsilon=0.013$. This exchange anisotropy mimics the effect of the
original single-ion anisotropy, resulting from the combination of crystal-field and
spin-orbit interactions. It is chosen such that the ordered AF state has spins in the $x$
direction, corresponding to the crystallographic a axis, as in experiment.
Following Ref.~\onlinecite{booth11} we neglect the anisotropy of the smaller coupling
$J'$ across a non-magnetic Co$^{3+}$ ion.

As discussed in the main text, for imperfect charge order, the classical ground state of
$\HSP$ is a cluster spin glass with well-defined short-range magnetic order, see also
Sec.~\ref{sec:glass} below. In Fig.~\ref{fig:xis}, we demonstrate the correlation between
$\xis$ and $\xic$. In particular, we obtain $\xispa\approx\xicpa$ for the system sizes
considered.


\section{Spin-wave calculation}

We calculate the excitation spectrum of $\HSP$ (which involves spatial disorder) using
real-space linear spin-wave theory\cite{bog86,mucci04,wessel05} on finite $L\times L$
lattices with periodic boundary conditions. This is done by expanding around a locally
stable classical state (or saddle point) of $\HSP$ as determined by the spin MC procedure
described in the main text.
The relevant classical states are in general non-collinear, with the
spin directions of the $\vec{S}_{i}$ characterized by angles $\theta_{i}$
and $\phi_{i}$. We start by introducing a new reference frame in
which $S_{i}^{z\prime}$ points locally in the direction of the magnetization
\begin{eqnarray*}
\left(\begin{array}{c}
S_{i}^{x}\\
S_{i}^{y}\\
S_{i}^{z}\end{array}\right) & = & \left(\begin{array}{ccc}
\mbox{cos}\theta_{i}\mbox{cos}\phi_{i} & -\mbox{sin}\phi_{i} & \mbox{sin}\theta_{i}\mbox{cos}\phi_{i}\\
\mbox{cos}\theta_{i}\mbox{sin}\phi_{i} & \mbox{cos}\phi_{i} & \mbox{sin}\theta_{i}\mbox{sin}\phi_{i}\\
-\mbox{sin}\theta_{i} & 0 & \mbox{cos}\theta_{i}\end{array}\right)\left(\begin{array}{c}
S_{i}^{x\prime}\\
S_{i}^{y\prime}\\
S_{i}^{z\prime}\end{array}\right)\end{eqnarray*}
 In this rotated frame $\vec{S}^{\prime}$, we apply the linearized
Holstein-Primakoff transformation: $S_{i}^{+\prime}=\sqrt{2S_{i}}a_{i}$,
$S_{i}^{-\prime}=\sqrt{2S_{i}}a_{i}^{\dagger}$, and $S_{i}^{z\prime}=S_{i}-a_{i}^{\dagger}a_{i}$,
where $S_{i}=1/2$ is the spin size at site $i$. The bilinear piece
$\mathcal{H}_{2}$ of the Hamiltonian then reads \begin{eqnarray}
\mathcal{H}_{2} & = & \sum_{ij}A_{ij}a_{i}^{\dagger}a_{j}+\frac{1}{2}\sum_{ij}\left(B_{ij}a_{i}^{\dagger}a_{j}^{\dagger}+h.c.\right),\label{eq:h_sw}\end{eqnarray}
 with the coefficients \begin{eqnarray}
A_{ij} & = & -2\delta_{ij}\sum_{k}F_{ik}^{zz}S_{k}+\sqrt{S_{i}}\sqrt{S_{j}}\left[F_{ij}^{xx}+F_{ij}^{yy}-\right.\nonumber \\
 & - & \left.i\left(F_{ij}^{xy}-F_{ji}^{xy}\right)\right],\label{eq:aij}\\
B_{ij} & = & \sqrt{S_{i}}\sqrt{S_{j}}\left[F_{ij}^{xx}-F_{ij}^{yy}+i\left(F_{ij}^{xy}+F_{ji}^{xy}\right)\right].\label{eq:bij}\end{eqnarray}
Here, the functions $F_{ij}^{\alpha\beta}$ read \begin{eqnarray*}
F_{ij}^{xx} & = & J_{ij}^{x}\mbox{cos}\theta_{i}\mbox{cos}\theta_{j}\mbox{cos}\phi_{i}\mbox{cos}\phi_{j}+\\
 & + & J_{ij}^{y}\mbox{cos}\theta_{i}\mbox{cos}\theta_{j}\mbox{sin}\phi_{i}\mbox{sin}\phi_{j}+J_{ij}^{z}\mbox{sin}\theta_{i}\mbox{sin}\theta_{j},\\
F_{ij}^{yy} & = & J_{ij}^{x}\mbox{sin}\phi_{i}\mbox{sin}\phi_{j}+J_{ij}^{y}\mbox{cos}\phi_{i}\mbox{cos}\phi_{j},\\
F_{ij}^{zz} & = & J_{ij}^{x}\mbox{sin}\theta_{i}\mbox{sin}\theta_{j}\mbox{cos}\phi_{i}\mbox{cos}\phi_{j}+\\
 & + & J_{ij}^{y}\mbox{sin}\theta_{i}\mbox{sin}\theta_{j}\mbox{sin}\phi_{i}\mbox{sin}\phi_{j}+J_{ij}^{z}\mbox{cos}\theta_{i}\mbox{cos}\theta_{j},\\
F_{ij}^{xy} & =- & J_{ij}^{x}\mbox{cos}\theta_{i}\mbox{cos}\phi_{i}\mbox{sin}\phi_{j}+J_{ij}^{y}\mbox{cos}\theta_{i}\mbox{sin}\phi_{i}\mbox{cos}\phi_{j}.\end{eqnarray*}

The bilinear Hamiltonian Eq.~\eqref{eq:h_sw} can be numerically
diagonalized using a bosonic Bogoliubov transformation,\cite{bog86,mucci04,wessel05}
and the resulting Hamiltonian reads (up to a constant) \begin{eqnarray}
\mathcal{H}_{2} & = & \sum_{\nu}\w_{\nu}b_{\nu}^{\dagger}b_{\nu},\label{eq:h_sw_diag}\end{eqnarray}
where $\w_{\nu}$ is the eigenfrequency associated to the $\nu$th
mode, and
\begin{eqnarray}
\left(\begin{array}{c}
a\\
a^{\dagger}\end{array}\right) & = & \left(\begin{array}{cc}
U & V\\
U^{\star} & V^{\star}\end{array}\right)\left(\begin{array}{c}
b\\
b^{\dagger}\end{array}\right),
\label{eq:bogol}
\end{eqnarray}
with $a^{T}=\left(a_{1},a_{2},\cdots,a_{\NS}\right)$, $b^{T}=\left(b_{1},b_{2},\cdots,b_{\NS}\right)$.
$U$ and $V$ are $\NS\times\NS$ matrices describing the transformation
to this diagonal basis.\cite{algo,zeromode}

Finally, we evaluate the rotationally averaged dynamic susceptibility,
$\chi''(\vec{q},\w)$, in the standard one-magnon approximation, see
Eq.~(2) of the main paper. For a single realization of disorder and
using the spin-wave transformation as described above we have
\begin{eqnarray}
\chi''(\vec{q},\w) & = & \sum_{\nu}\sum_{\alpha}\left|M_{\nu}^{\alpha}\left(\vec{q}\right)\right|^{2}\delta\left(\w-\w_{\nu}\right).
\label{eq:susc_final}
\end{eqnarray}
Here $\alpha=x,\, y,\, z$ and $M_{\nu}^{\alpha}\left(\vec{q}\right)$
represents the matrix element $\langle0|S^{\alpha}\left(\vec{q}\right)|\nu\rangle$
of the Fourier-transformed spin operator with the eigenmode $|\nu\rangle$:
\begin{eqnarray}
M_{\nu}^{\alpha}\left(\vec{q}\right) & = & \frac{1}{\sqrt{L^{2}}}\sum_{i}e^{-i\vec{q}\cdot\vec{r}_{i}}M_{\nu}^{\alpha}\left(\vec{r}_{i}\right),
\label{eq:ftm}
\end{eqnarray}
 where \begin{eqnarray}
M_{\nu}^{x}\left(\vec{r}_{i}\right) & = & \sqrt{\frac{S_{i}}{2}}\left[U_{i\nu}\left(\mbox{cos}\theta_{i}\mbox{cos}\phi_{i}+i\mbox{sin}\phi_{i}\right)+\right.\nonumber \\
 & + & \left.V_{i\nu}^{\star}\left(\mbox{cos}\theta_{i}\mbox{cos}\phi_{i}-i\mbox{sin}\phi_{i}\right)\right],\label{eq:mx}\\
M_{\nu}^{y}\left(\vec{r}_{i}\right) & = & \sqrt{\frac{S_{i}}{2}}\left[U_{i\nu}\left(\mbox{cos}\theta_{i}\mbox{sin}\phi_{i}-i\mbox{cos}\phi_{i}\right)+\right.\nonumber \\
 & + & \left.V_{i\nu}^{\star}\left(\mbox{cos}\theta_{i}\mbox{sin}\phi_{i}+i\mbox{cos}\phi_{i}\right)\right],\label{eq:my}\\
M_{\nu}^{z}\left(\vec{r}_{i}\right) & = & \sqrt{\frac{S_{i}}{2}}\left[-\mbox{sin}\theta_{i}\left(U_{i\nu}+V_{i\nu}^{\star}\right)\right].\label{eq:mz}
\end{eqnarray}
As momentum is not a good quantum number in a disordered system, the matrix element
$M_{\nu}^{\alpha}\left(\vec{q}\right)$, for any given $\vec{q}$, will in general be
non-zero for all eigenmodes $\nu$.


\section{Charge disorder and hourglass spectrum: Additional results}

\begin{figure*}
\includegraphics[width=1\textwidth]{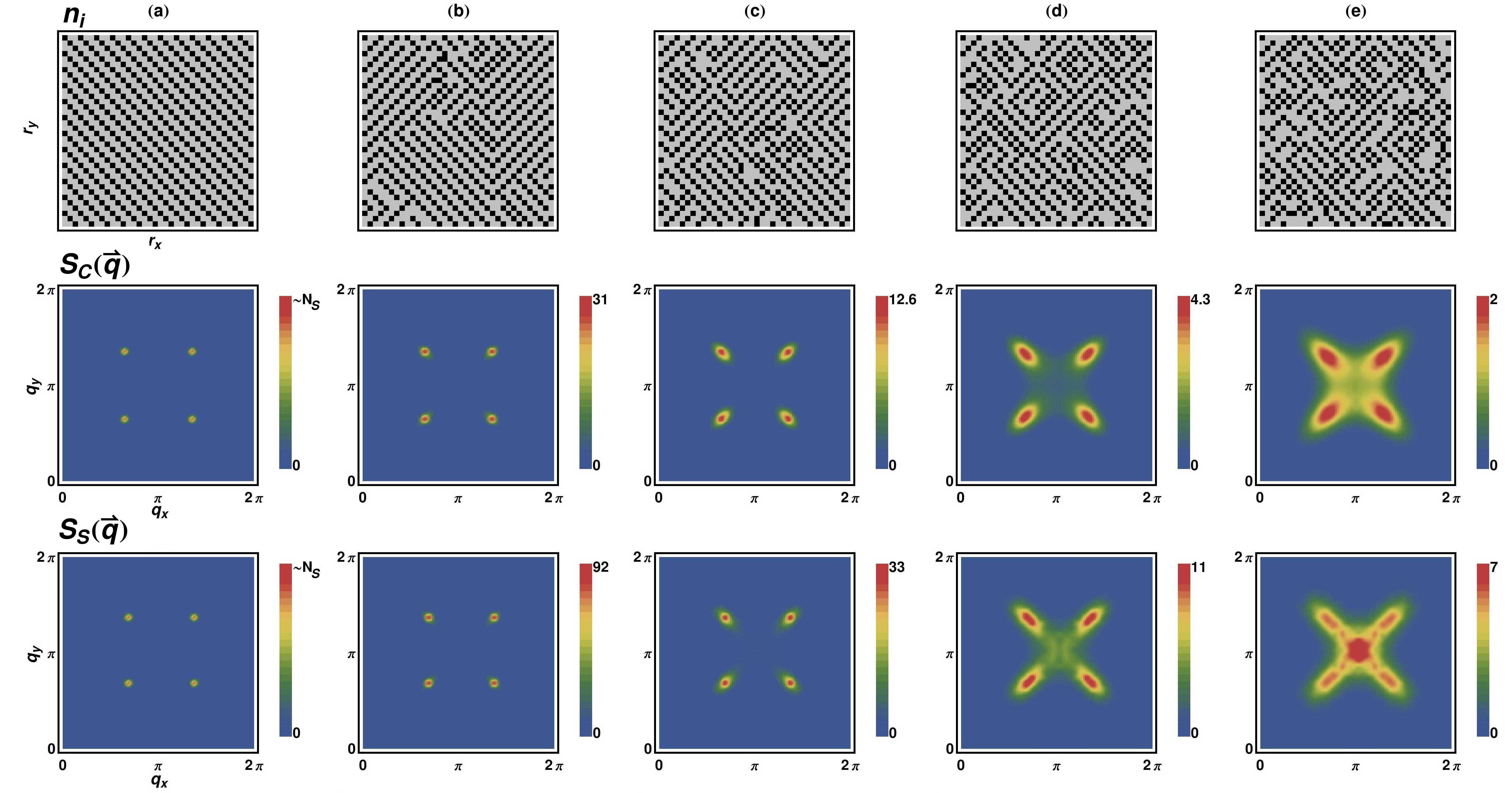}
\caption{\label{fig:all1}
Numerical results for individual charge configurations
$n_{i}$ (top), the charge structure factor $\Sc(\vec{q})$ (middle),
and the spin structure factor $\Ss(\vec{q})$ (bottom) for disordered
stripes with $L=36$. From (a) to (e), the degree of charge disorder increases
as in Fig.~2 of the main paper. For the charge sector, the parameters
are $V_{1}=4$, $V_{2}=-2$, $V_{3}=2$, $V_{4,5}=0$, $T=1$, and
(a) perfect stripes, (b) $\NMC=2\times10^{6}$, $\xicpa\simeq19$,
$\xicpe\simeq8$, (c) $\NMC=0.5\times10^{6}$, $\xicpa\simeq10$,
$\xicpe\simeq4$, (d) $\NMC=0.1\times10^{6}$, $\xicpa\simeq5$, $\xicpe\simeq2$,
(e) $\NMC=0.025\times10^{6}$, $\xicpa\simeq3$, $\xicpe\lesssim1$.
The spin excitation spectra corresponding to the situations in panels
(a) to (e) are displayed in Fig.~2 of the main paper, which also
specifies the parameters of the spin model $\HSP$.
}
\end{figure*}

The evolution of the spin excitations with increasing charge disorder
is shown in Fig.~2 of the main paper. Here we complement this by
Fig.~\ref{fig:all1} which displays snapshots of the charge configuration
as well as $\Sc(\vec{q})$ and $\Ss(\vec{q})$ for the same parameters
as in Fig.~2. The figure reveals a very similar progressive and anisotropic
broadening in $\Sc(\vec{q})$ and $\Ss(\vec{q})$ for weak and moderate
disorder. This is different for strong disorder, Fig.~\ref{fig:all1}(e),
where the IC signal in $\Ss(\vec{q})$ has almost disappeared in favor
of a broad peak at the AF wavevector $(\pi,\pi)$, while $\Sc(\vec{q})$
still displays a clear IC structure. This corresponds to
a state where disorder-induced frustration has destroyed the period-3
spin correlations, leaving behind weakly correlated local AF clusters
-- this is further illustrated in Fig.~\ref{fig:spins} below.

\begin{figure}[!b]
\includegraphics[width=0.45\textwidth]{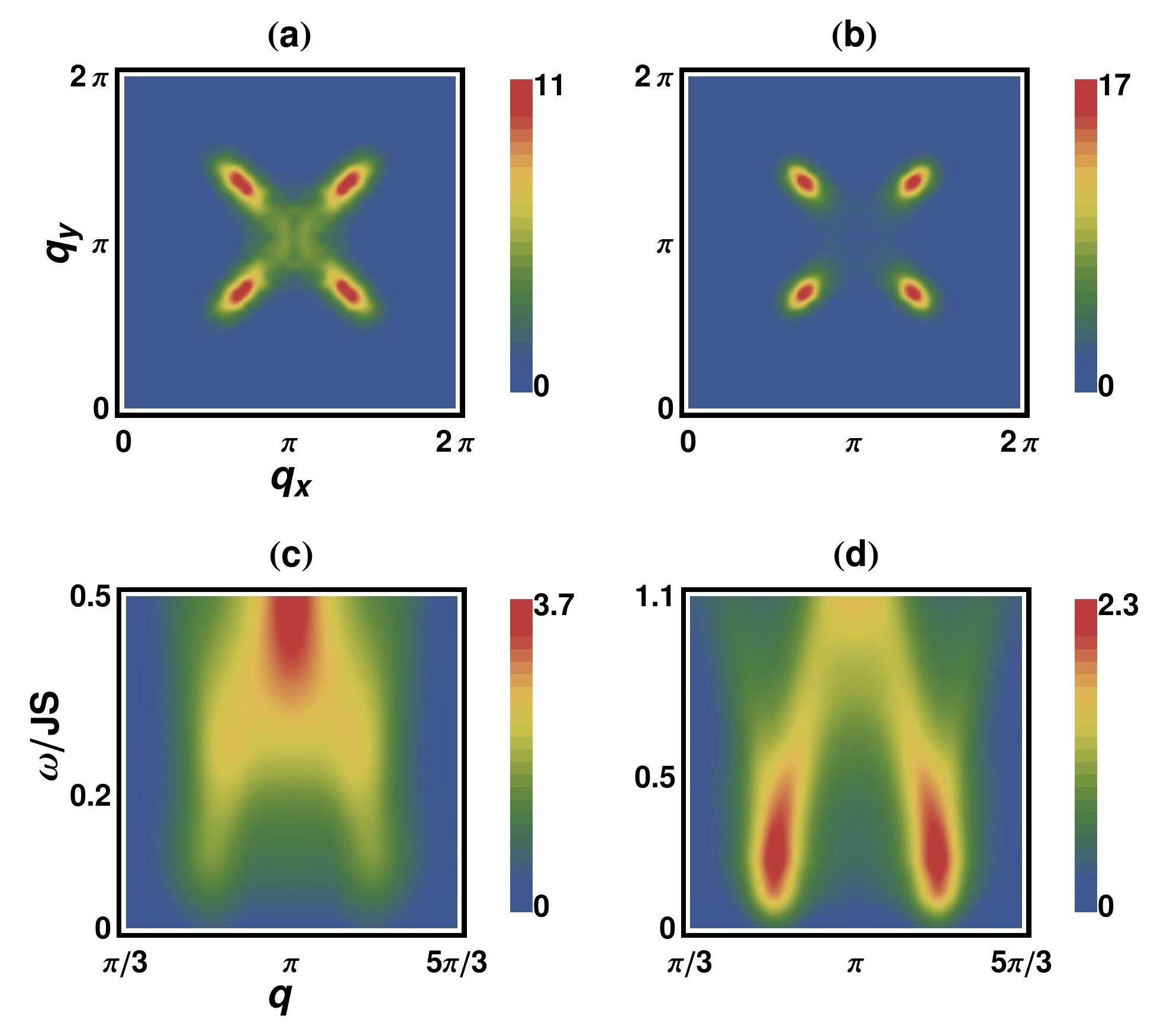}
\caption{
\label{fig:jjcomp}
Influence of the coupling ratio $J/J'$ on the
magnetism, for charge-stripe parameters as in Fig.~2(d) of the main paper.
(a,b) $\Ss(\vec{q})$ and (c,d) low-energy part of $\chi''(\vec{q},\w)$
along $\vec{q}=(q,q)$ for (a,c) $J^{\prime}=0.05J$ and (b,d) $J^{\prime}=0.15J$.
}
\end{figure}

\subsection{Spin-sector parameters}

We have performed calculations for disordered stripes with varying
$J/J'$, with sample results in Fig.~\ref{fig:jjcomp}. These confirm
that $J/J'$ decisively influences the form of the magnetic spectrum:
While an hourglass emerges for $J/J'=20$, for $J/J'\sim7$
there is neither a large-intensity feature at $(\pi,\pi)$ nor a
{}``vertical'' dispersion. Instead, large weight is concentrated
at low energies close to $\Qs$, even though the outward dispersing
branches are weakened by disorder. Moreover, the structure factor
$\Ss(\vec{q})$ reveals that the spin sector is less sensitive to
charge disorder for smaller $J/J'$: stronger inter-stripe coupling
stabilizes the IC state against glassy behavior.

We note that $J/J'$ is often deduced via the experimentally measured
location of the saddle point, $\wR/(JS)$. Our results in Fig.~2 of the main paper
show that care is required, because $\wR/(JS)$ also depends on the degree
of stripe disorder.

\subsection{Charge-sector parameters}

\begin{figure}[t]
\includegraphics[width=0.45\textwidth]{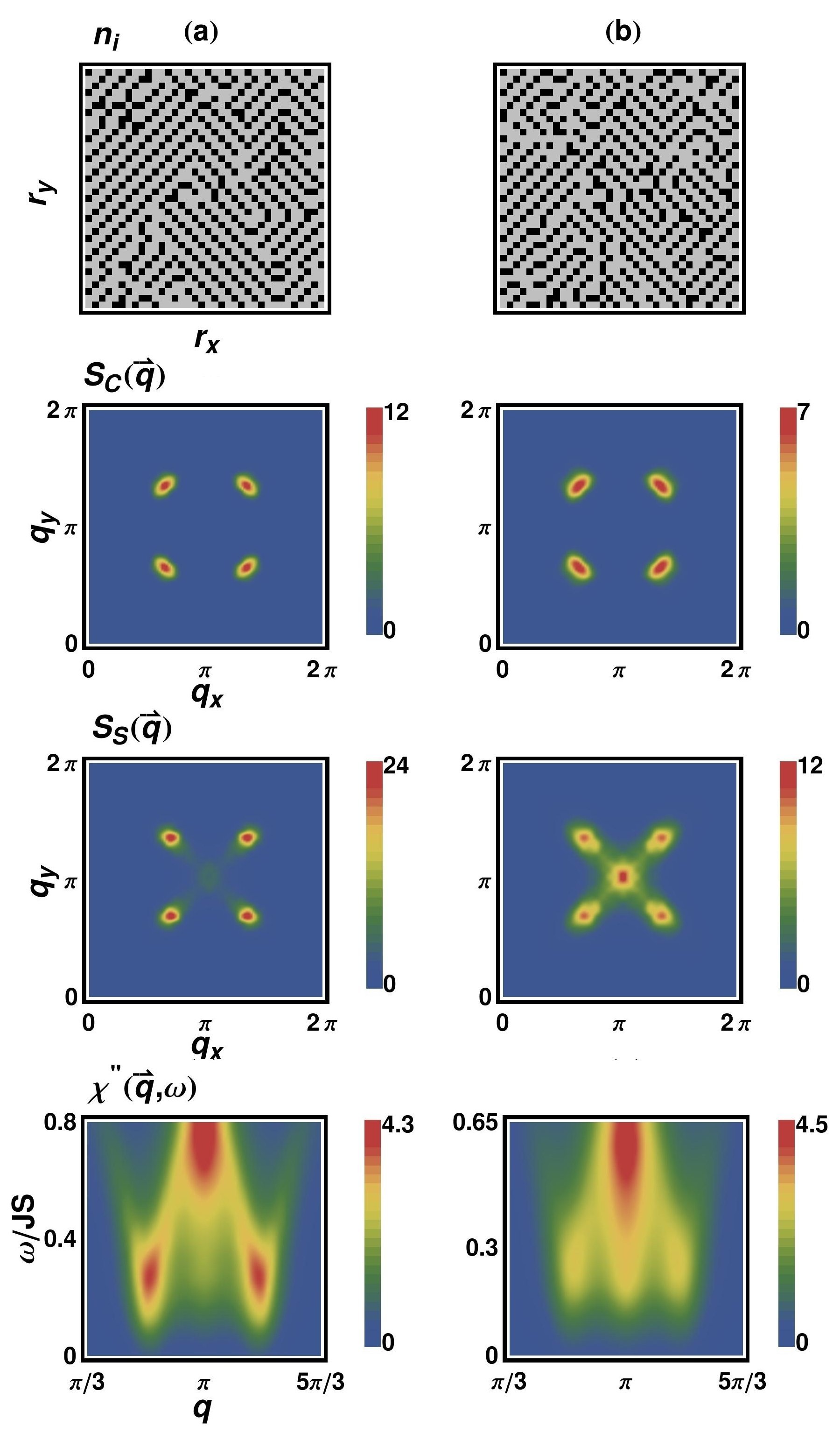}
\caption{\label{fig:all2}
Numerical results for disordered stripes as in Fig.~\ref{fig:all1},
but here for stripe parameters with $\xicpa<\xicpe$. From top to
bottom: charge configuration $n_{i}$, charge structure factor $\Sc(\vec{q})$,
spin structure factor $\Ss(\vec{q})$, and the low-energy part of
the magnetic excitations, $\chi''(\vec{q},\w)$, along $\vec{q}=(q,q)$.
The parameters for the charge sector are $L=36$, $V_{1}=2$, $V_{2}=0$,
$V_{3}=2$, $V_{4}=-2,V_{5}=0.5$, $T=1$, and (a) $\NMC=4\times10^{4}$,
$\xicpa\simeq5$, $\xicpe\simeq10$, (b) $\NMC=0.5\times10^{4}$,
$\xicpa\simeq2$, $\xicpe\simeq5$. For the spin sector we have $J/J'=20$
and the $\delta$ peaks in $\chi''(\vec{q},\w)$ were replaced by
Lorentzians with width $\Gamma=0.1JS$. Remarkably, $\Ss(\vec{q})$
displays anisotropic broadening with $\xispa>\xispe$, despite $\xicpa<\xicpe$
in the charge sector. Consequently, the spin excitations here are
qualitatively similar to that in Fig.~2 of the main paper (for a
comparable degree of charge disorder).
}
\end{figure}

As mentioned in Sec.~\ref{sec:charge}, the disordered stripes obtained
from our MC procedure quite generically display $\xicpa>\xicpe$ --
this applies to all cases in Fig.~\ref{fig:all1}. Results for the
opposite situation, $\xicpa<\xicpe$, are shown in Fig.~\ref{fig:all2}
-- those were obtained using $V_{4,5}\neq0$ in $\mathcal{H}_{C}$,
Eq.~\eqref{eq:h_charge}. Remarkably, even for this kind of disorder
configuration we find $\xispa>\xispe$ which thus appears universal.
This can be rationalized considering that we work in the limit $J\gg J'$:
Here, intact AF domains will show nearly perfect order, while correlations
between AF domains, mediated by the small $J'$, can easily be destroyed
by spatial disorder. As the geometry implies that $J$ stabilizes
correlations {\em along} the stripe direction, this explains $\xispa>\xispe$.
It is then not surprising that the evolution of the spin excitation
spectrum in the case $\xicpa<\xicpe$, shown in the bottom row of
Fig.~\ref{fig:all2}, in qualitatively similar to that in the case
$\xicpa>\xicpe$, Fig.~2 of the main paper: While both high-intensity
features at $\Qs$ and outward dispersing branches at low energy are
still visible for weak disorder, those tend to be suppressed for strong
disorder, where instead an apparent {}``vertical'' dispersion appears
at intermediate energies.

An interesting feature in Fig.~\ref{fig:all2}(b) is a considerable
enhancement of the intensity at $\left(\pi,\pi\right)$ even for energies
well below the saddle point, coexisting with low-energy weight near $\Qs$.
This can be traced back to the momentum-space structure of $\Ss\left(\vec{q}\right)$,
which shows peaks both at $\Qs$ and $(\pi,\pi)$, arising from the
coexistence of relatively well-defined period-3 structures and small
AF clusters. (We have not observed this pronounced coexistence for
disordered stripes with $\xicpa>\xicpe$, implying that microscopic
details of the charge sector play some role here.) A further increase
of the charge disorder tends to transfer spectral weight from $\Qs$
to $(\pi,\pi)$, similar to Fig.~2(e) of the main paper.

%
%


\section{Comparison to broadened perfect-stripe spectra}

In order to further illustrate the importance of a proper treatment of stripe disorder,
we provide a comparison of our results to those from spin-wave calculations
for perfectly ordered stripes, where disorder is simply accounted for by an ad-hoc broadening in
momentum space. This procedure has been employed previously in the literature: For
instance, in Ref.~\onlinecite{booth11} reasonable agreement with the experimental
data concerning the intensity distribution in momentum space was achieved.

\begin{figure}[t]
\includegraphics[width=0.45\textwidth]{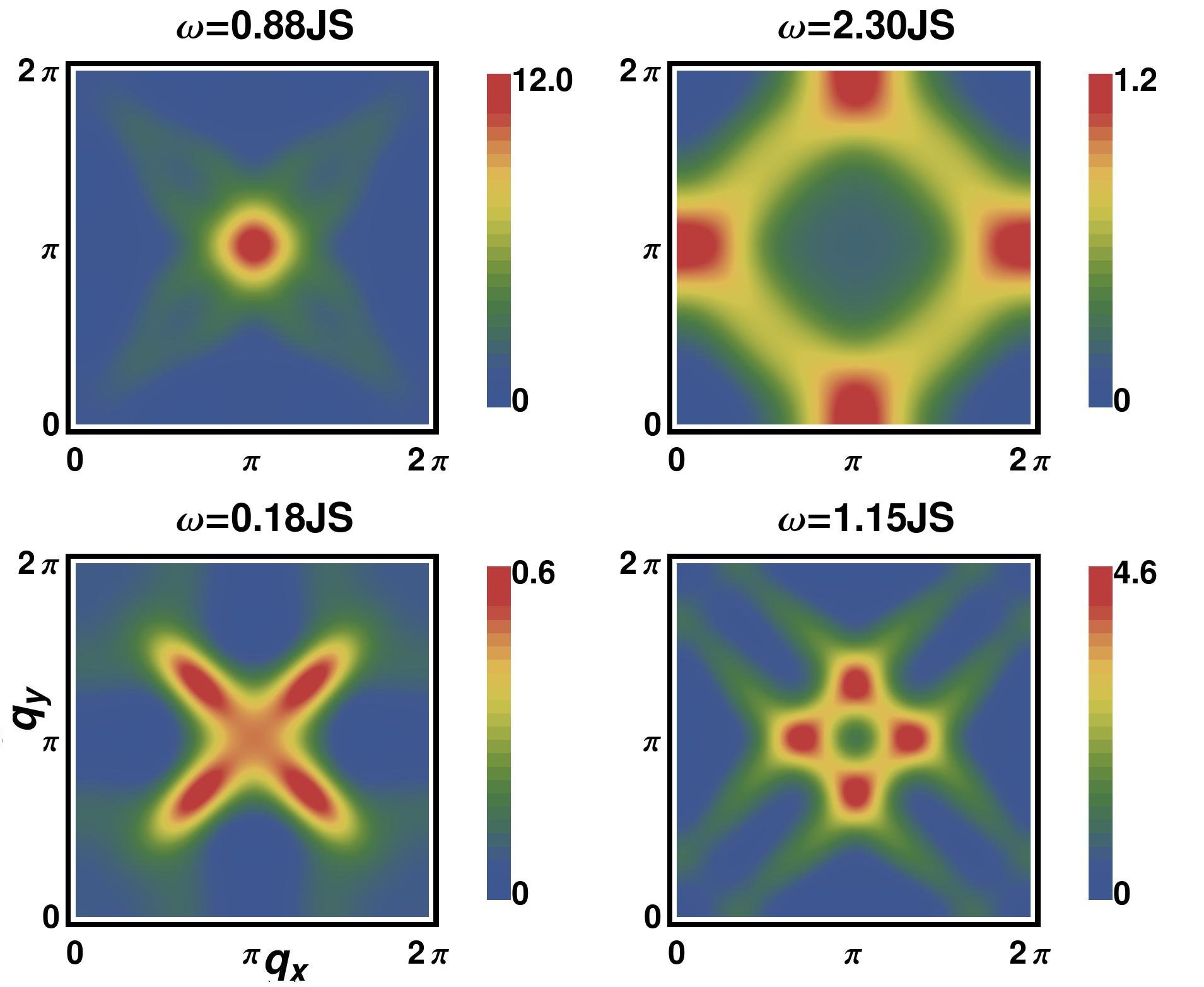}
\caption{\label{fig:swconv}
Constant-energy cuts of $\chi''(\vec{q},\w)$ as a function of momentum $\vec{q}$ as in
Fig.~3 of the main paper, but now for perfect stripes with $J/J'=20$ and an anisotropic
ad-hoc broadening in momentum space of $\gamma_\perp = 0.18\pi$ and $\gamma_\parallel =
0.09\pi$. These results are broadly similar to that in Fig.~4e-h of
Ref.~\onlinecite{booth11}, where, however, the full crystal-field model for the Co$^{2+}$
ions (instead of an effective spin-1/2 model) was used.
}
\end{figure}

We have implemented the ad-hoc broadening for the spin-1/2 model in Eq.~\eqref{eq:h_spin}
following Ref.~\onlinecite{booth11}, by convolving $\chi''(\vec{q},\w)$ at a fixed
energy $\w$ with a two-dimensional anisotropic Gaussian. This introduces two fit
parameters -- the momentum-space widths $\gamma_\parallel$ and $\gamma_\perp$ -- which
mimick stripe disorder and thus replace the physical parameter $\xic$ of our calculation.
Results for broadened perfect-stripe spectra with $J/J'=20$
are shown in Fig.~\ref{fig:swconv}, where $\gamma_\parallel$ and $\gamma_\perp$ were
chosen as to match the momentum width of the experimentally observed excitations at low
energy (3\,meV in Fig.~4 of Ref.~\onlinecite{booth11}). (The same parameters are used for
the dashed curves in Fig.~4 of the main text.)
As a result, the momentum-space intensity distributions are roughly similar
to that obtained from the full calculation displayed in Fig.~3 of the main paper (and also
roughly similar to that of the experimental data in Fig.~4 of Ref.~\onlinecite{booth11}).

\begin{figure}[t]
\includegraphics[width=0.45\textwidth]{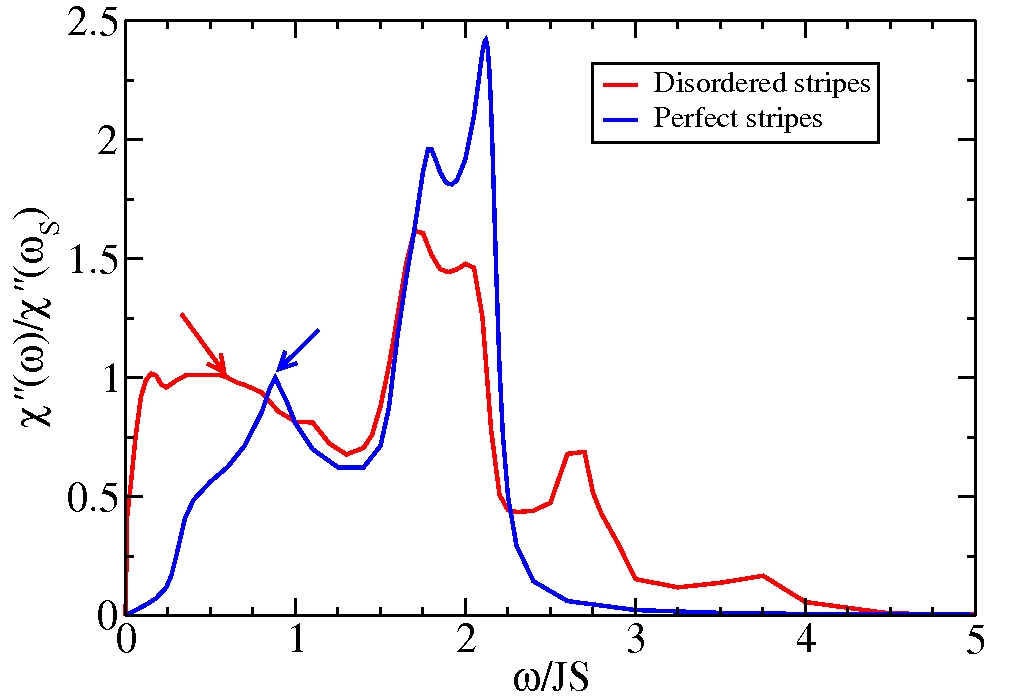}
\caption{\label{fig:kint}
Momentum-integrated spectral weight as function of energy $\w$, comparing the result from
a full simulation of disordered stripes (with $\xicpa\simeq5$, $\xicpe\simeq2$ as in
Fig.~2d of the main paper) to that of perfect stripes (additional momentum-space
broadening does not influence this curve). In both cases $J/J'=20$, and a Lorentzian
energy broadening of $\Gamma=0.1JS$ has been employed. The arrows denote the energy
position $\wR$ of the $(\pi,\pi)$ saddle point, and the intensities have been rescaled
to match at $\wR$.
The intensity peaks in the range $1.7-2.1 JS$ originate primarily from the upper of the
two spin-wave branches.\cite{booth11}
}
\end{figure}

However, a closer inspection at the energy dependence of the intensities reveals strong
differences. This is already seen in the comparison at selected energies as shown in Fig.~4 of
the main paper, and becomes clear when considering the momentum-integrated intensity
$\chi''(\w)$, Fig.~\ref{fig:kint}. While the perfect-stripe-based result has rather
little intensity at low energies and a sharp saddle-point feature at $\w_S$, the result
for disordered stripes has large low-energy weight (note that even the anisotropy gap is
almost completely smeared) and a broad saddle-point feature.
As a result, the latter calculations yields only moderate intensity variations between
$0.1\,JS$ and $2.5\,JS$, in accord with experiment (compare Fig.~4 of the main paper). In
contrast, much larger intensity variations with energy arise from perfect stripes.
In addition, Fig.~\ref{fig:kint} underlines the downward renormalization of $\wR$ and also
the overall increase of the bandwidth caused by disorder. Therefore, extracting $JS$ (or
$J/J'$) by matching theoretical and experimental features of the spectra (like the
saddle-point location $\wR$) requires care.

\section{Spin-glass signatures}
\label{sec:glass}

\begin{figure}[t]
\includegraphics[width=0.45\textwidth]{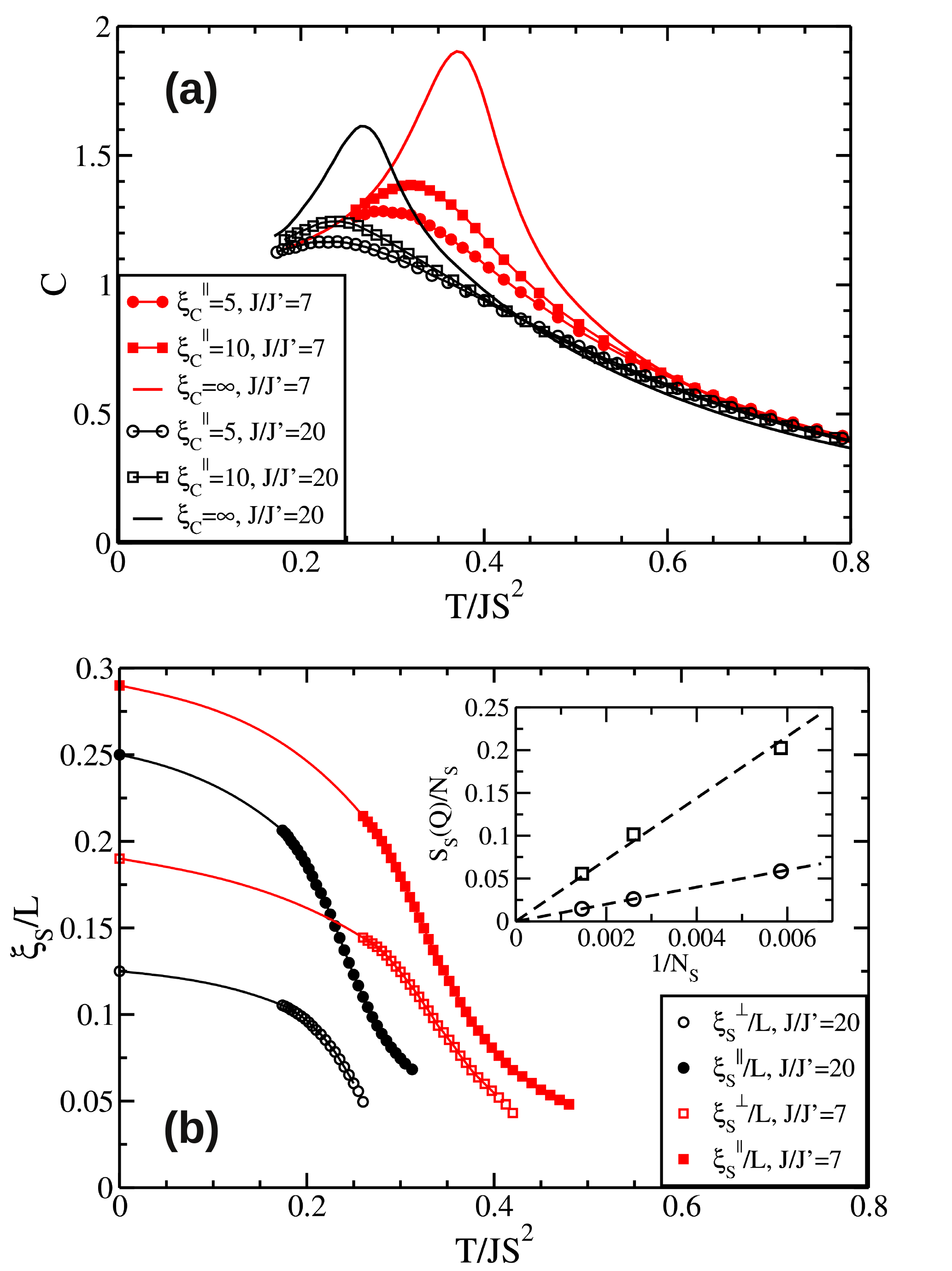}
\caption{\label{fig:cv}
Characterization of glassy behavior in the spin sector,
as obtained from the MC simulations of $\HSP$ for $L=36$. (a) Specific
heat $C$ as function of temperature, for different $J/J'$ (see legend)
and different amounts of charge disorder: circles: $\xicpa\approx5$,
$\xicpe\approx2$, squares: $\xicpa\approx10$, $\xicpe\approx5$,
solid line: perfect stripes. (b) Spin correlation lengths $\xispa$,
$\xispe$ as function of temperature for different $J/J'$ and $\xicpa\approx5$,
$\xicpe\approx2$. Here, the results at $T=0$ have been obtained
from locally stable classical states. Error bars are typically smaller
than the symbol size.
Note that, for higher $T$ (where no data points are shown), $\Ss(\vec{q})$ does not show
well-defined peaks at $\Qs$ but instead a broad maximum at $(\pi,\pi)$, i.e., we observe
a temperature-dependent onset of incommensurability.
Inset: $T=0$ spin structure factor $\Ss(\vec{q})$
at one of the ordering wavevectors $\Qs$ normalized by the number
of spins $\NS$ as a function of $1/\NS$. For both curves we have
$J/J'=20$ and different amounts of charge disorder: circles:
$\xicpa\approx5$, $\xicpe\approx2$, squares: $\xicpa\approx10$,
$\xicpe\approx5$.
}
\end{figure}

\begin{figure*}[t]
\includegraphics[width=1\textwidth]{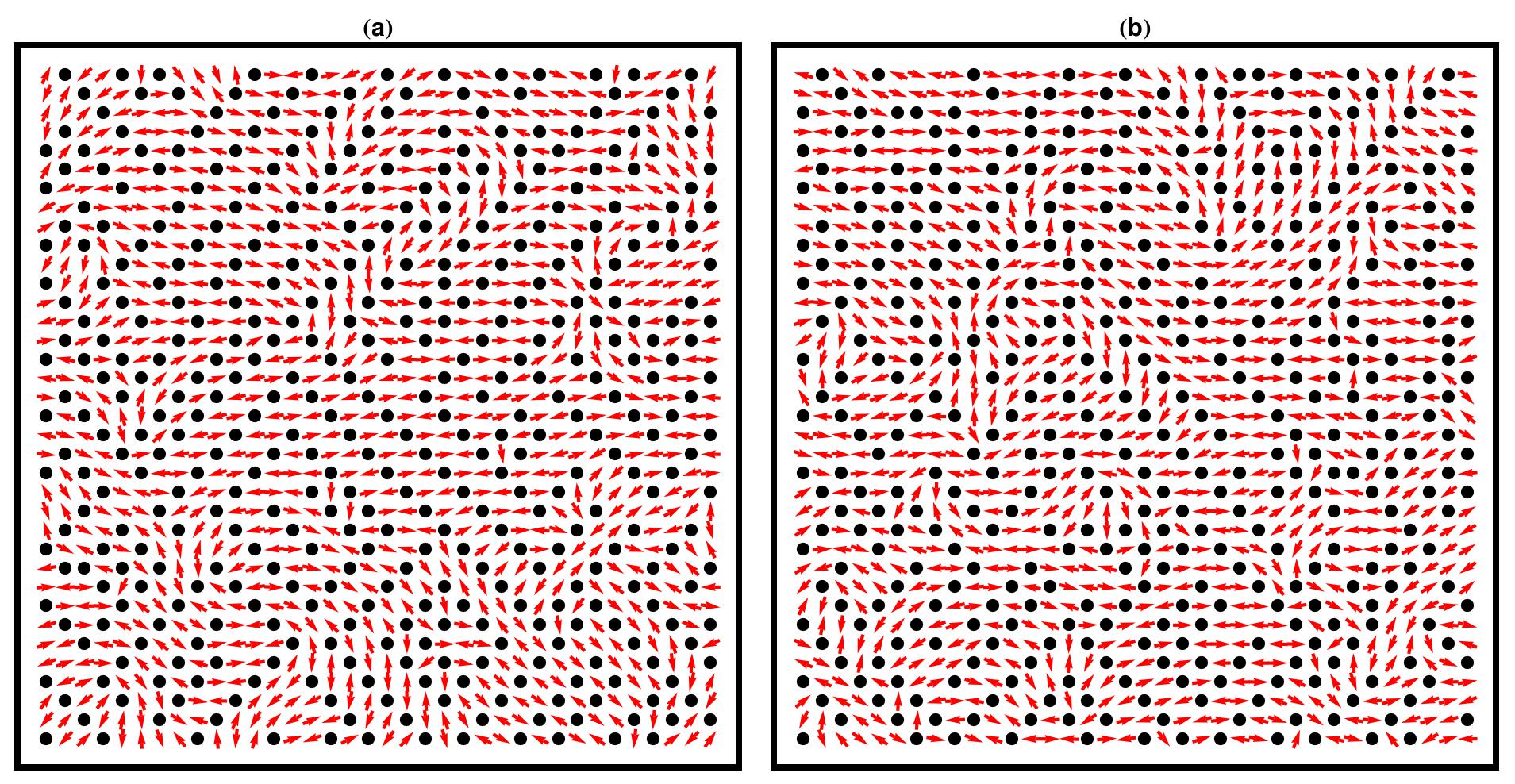}
\caption{\label{fig:spins}
Real-space spin configurations in classical ground
states of $\HSP$ with $L=36$, for the same parameters and charge
configurations as in Fig.~\ref{fig:all1}(d) ($\xicpa\simeq5$, $\xicpe\simeq2$
(a)) and Fig.~\ref{fig:all1}(e) ($\xicpa\simeq3$, $\xicpe\lesssim1$
(b)). The dots show the non-magnetic sites, while the arrows show
the $x$ and $y$ components of the $\vec{S}_{i}$ (the $z$ components
are tiny due to the strong anisotropy caused by $\delta$).
}
\end{figure*}

We can characterize spin-glass behavior in our MC simulations for
the spin sector by monitoring the specific heat $C$ and the magnetic
correlation length $\xis$ as function of temperature $T$, both shown
in Fig.~\ref{fig:cv}. Panel (a) shows that the specific-heat peak,
which occurs near the mean-field N\'{e}el temperature in the perfect-stripe
case, is strongly broadened and shifted to a lower $T_{f}$ with increasing
stripe disorder. This effect is more pronounced for weakly coupled
stripes, i.e., large $J/J'$. As can be seen in panel (b), the correlation
length $\xis$ grows upon cooling, in particular in the temperature
regime near $T_{f}$, but remains much smaller than the system size
as $T\to0$. Again, this is more pronounced for large $J/J'$, which
also causes a larger anisotropy as measured by the ratio $\xispa/\xispe$.

In addition, we have also calculated the spin-glass correlation length $\xiSG$ from the
correlations of the Edwards-Anderson order parameter.\cite{mc_ea,av12} This correlation
length strongly increases below $T_f$ as well and reaches values of the order of the
system size as $T\to 0$ (not shown). Importantly, $\xiSG$ is always larger than $\xis$
(except for very large $T$).

All these results are consistent with the behavior of a spin glass with
short-range magnetic order and a mean-field freezing temperature $T_{f}$.
Clearly, glassiness increases with increasing charge disorder and
with increasing $J/J'$, as indicated by a broader freezing regime
and smaller $\xis$. For large $J/J'$, the glassy state consists
of ordered AF clusters (formed by intact stripe segments of Co$^{2}+$
ions) which are weakly correlated among each other (see also Fig.~\ref{fig:spins})
-- this justifies to call this a cluster spin glass. Note that the
ground state is expected to remain glassy even in the limit of $J\sim J'$,
where the strongly disordered system corresponds to a vacancy-doped
frustrated spiral magnet which displays more conventional spin-glass
(instead of cluster-spin-glass) behavior.

We note that, in our two-dimensional spin model $\HSP$, a true spin-glass
transition does not exist.\cite{fischer} Therefore, we refrain
from a detailed finite-size scaling analysis of the thermodynamic
and magnetic properties \cite{mc_ea,av12} of $\HSP$ -- such an analysis
would also be complicated by the fact that we have no straightforward
way to introduce precisely the same amount of microscopic disorder
at different $L$ due to the use of a non-equilibrium MC protocol.
We have, however, investigated the $L$ dependence of our results
for the spin structure factor $\Ss(\vec{q})$ at fixed $\xic$ (which
is indeed a very good estimator of disorder). For $\Ss(\vec{q})$
we essentially find no dependence on the system size in the range
$L=24$--$48$. In particular, $\Ss(\Qs)/\NS\rightarrow0$
as $1/\NS\rightarrow0$ (see the inset of Fig. \ref{fig:cv}(b)) indicating
once more static short-range order with a vanishing magnetic order
parameter. Similarly, the susceptibility $\chi''(\vec{q},\w)$ displays
no signs of size dependence (not shown).

Finally, we illustrate the cluster-spin-glass state by showing specific
saddle points of $\HSP$ in Fig.~\ref{fig:spins}: these are portraits
of the spin configurations of locally stable classical states. The
parameters correspond to moderately and strongly disordered stripes
as in Fig.~2(d,e) and Fig.~\ref{fig:all1}(d,e). The configurations
nicely show clusters with local AF order and essentially random relative
orientations. In particular, the states are non-collinear, i.e., the
small easy-axis anisotropy $\epsilon$ is less effective in the presence
of disorder.
